\documentclass[prx,aps,twocolumn,nopacs,superscriptaddress,nofootinbib]{revtex4-2}

\usepackage{amsmath}
\usepackage{amssymb}
\usepackage{amsfonts}
\usepackage{bm} 
\usepackage{bbm}
\usepackage{braket}
\usepackage{color}
\usepackage{comment}
\usepackage{dcolumn} 
\usepackage{dsfont}
\usepackage{enumerate}
\usepackage{epsfig}
\usepackage{esint}
\usepackage[T1]{fontenc}
\usepackage{framed}
\usepackage{gensymb}
\usepackage{graphicx} 
\usepackage[colorlinks,linkcolor=blue,citecolor=blue,urlcolor=blue,hyperindex,driverfallback=dvipdfm]{hyperref}
\usepackage{indentfirst}
\usepackage{lmodern}
\usepackage{mathrsfs}
\usepackage{mathtools}
\usepackage{multirow}
\usepackage{psfrag}
\usepackage{pst-all}
\usepackage{soul}
\usepackage{xcolor}
\usepackage{xspace}
\usepackage{orcidlink}

\newcommand{\abs}[1] {\mathopen{}\left|#1\right|\mathclose{}}

\newcommand{\ccpar}[1] {\mathopen{}\left(#1\right)\mathclose{}}
\newcommand{\sqpar}[1] {\mathopen{}\left[#1\right]\mathclose{}}
\newcommand{\clpar}[1] {\mathopen{}\left\{#1\right\}\mathclose{}}

\def\ii{{\rm i}}  \def\ee{{\rm e}}
  \def\kB{{k_{\rm B}}}
\def\Ree{{\rm Re}}  \def\Imm{{\rm Im}}
\newcommand{\pd}[2] {\mathopen{}\frac{\partial#1}{\partial#2}\mathclose{}}

\def\rb{{\bf r}}  \def\Rb{{\bf R}}
  
\def\xb{{\bf x}} \def\yb{{\bf y}}  \def\zb{{\bf z}}
\def\xx{\hat{\bf x}}  \def\yy{\hat{\bf y}}  \def\zz{\hat{\bf z}}
  
\def\rr{\hat{\bf r}}    
\def\kb{{\bf k}}    
       \def\Bb{{\bf B}}
\def\Kb{{\bf K}} \def\eb{{\bf e}}
\def\Eb{{\bf E}}  \def\Db{{\bf D}}  
\def\Bb{{\bf B}}  \def\Hb{{\bf H}}    
\def\Jb{{\bf J}}  \def\pb{{\bf p}}
\def\nb{{\bf n}}
\def\vF{v_{\rm F}}    \def\EF{{E_{\rm F}}}  
\def\NF{{N_{\rm F}}}
\def\eps{\epsilon}  \def\vep{\varepsilon}  \def\ww{\omega}
\def\Am{\mathcal{A}}  \def\Bm{\mathcal{B}}  \def\Cm{\mathcal{C}}  \def\Dm{\mathcal{D}}
\def\Gm{\mathcal{G}}  
  \def\Lm{\mathcal{L}}  
  \def\Vm{\mathcal{V}}  
\def\Mm{\mathcal{M}}  
    \def\epseff{\epsilon_{\rm eff}}
  \def\phiext{\phi^{\rm ext}}  
  
\def\rhoind{\rho^{\rm ind}}

\begin{document}

\title{Chiral near-field control of quantum light generation using magneto-optical graphene}

\author{Mikkel~Have~Eriksen\,\orcidlink{0000-0002-0159-0896}}
\affiliation{POLIMA---Center for Polariton-driven Light--Matter Interactions, University of Southern Denmark, Campusvej 55, DK-5230 Odense M, Denmark}

\author{Joel~D.~Cox\,\orcidlink{0000-0002-5954-6038}}
\email[Joel~D.~Cox: ]{cox@mci.sdu.dk}
\affiliation{POLIMA---Center for Polariton-driven Light--Matter Interactions, University of Southern Denmark, Campusvej 55, DK-5230 Odense M, Denmark}
\affiliation{Danish Institute for Advanced Study, University of Southern Denmark, Campusvej 55, DK-5230 Odense M, Denmark}

\begin{abstract}
We theoretically explore strategies to actively control photon emission from quantum light sources by leveraging the large magneto-optical response of graphene. The quantum electrodynamic response of graphene---characterized by the Purcell factor and the Lamb shift of a proximal emitter---is analyzed for extended two-dimensional sheets, one-dimensional nanoribbons, and zero-dimensional nanodisks, all of which are endowed with an intrinsic chiral near-field response under a static perpendicular magnetic field. Using rigorous semianalytical models of these systems, we reveal that the emission properties can be readily tuned by variations in doping charge carrier density and applied magnetic field strength, both with respect to magnetoplasmon resonances (at infrared frequencies) and Shubnikov-de-Haas oscillations (entering telecommunication bands) associated with optical transitions between discrete Landau levels. Localized magnetoplasmons in graphene nanoribbons are predicted to induce large dissymmetry in the spontaneous emission from left-hand and right-hand circularly polarized transitions in a proximal quantum emitter, presenting applications for chiral quantum optical waveguiding. This chiral dissymmetry is further enhanced in gyrotropic graphene nanodisks, signaling that the spatial shaping of near-fields in nanostructured graphene can significantly boost the intrinsic chiral response induced by the magnetic field. These results indicate that magneto-optical graphene constitutes a versatile and highly tunable platform for quantum light generation and manipulation at the nanoscale.
\end{abstract}

\date{\today}
\maketitle

\section{Introduction}

Photon generation is of key importance not only for fundamental explorations in quantum physics~\cite{couteau2023applications2}, but also in a plethora of light-based technological applications, such as super-resolution imaging~\cite{koenderink2022super}, quantum cryptography~\cite{bozzio2022enhancing}, and quantum information processing~\cite{couteau2023applications}. While the numerous available quantum light sources---from natural atoms and molecules to engineered quantum dots and optically active defects---exhibit diverse photon emission rates and operational wavelengths~\cite{aharonovich2016solid,toninelli2021single}, their properties are not easily controlled in an active manner. Fortunately, the spectral and temporal dynamics of quantum light emitters (QEs), along with the polarization state and directionality of the emitted photons, depend crucially on their electromagnetic environment and the associated local photonic density of states (LDOS)~\cite{novotny2012principles,barnes2020classical}. The interaction of QEs with engineered optical resonators thus enables modification of their spontaneous emission rate and transition energy through the Purcell effect and Lamb shift, respectively~\cite{hohenester2020nano,verlekar2025giant}, while the LDOS can be tailored in nanophotonic environments to cause preferential generation of photons with certain polarization and propagation characteristics \cite{lodahl2017chiral}. However, the development of integrated nanophotonic devices that enable active tuning of quantum light generation still presents a fundamental challenge.

The relatively strong light-matter interaction and optoelectronic tunability of graphene renders the carbon monolayer an appealing material platform for integrated photonics~\cite{bonaccorso2010graphene}. In electrically doped graphene, the optical response is dominated by collective oscillations of conduction electrons, i.e., plasmons, at frequencies $\hbar\omega\lesssim\EF$, while interband transitions $\hbar\omega>2\EF$ lead to broadband $\sim2.3\%$ optical absorption~\cite{gonccalves2016introduction}. For achievable doping levels $\EF\sim1$\,eV, graphene plasmons can thus intensify impinging electromagnetic fields in the terahertz (THz) and infrared (IR) spectral regimes~\cite{garciadeabajo2014graphene}, while the absorption of visible light is impeded by Pauli blocking below the $\hbar\omega\simeq 2\EF$ threshold \cite{nair2008fine}. Depending on their relevant dipole-active transition energies, either of these mechanisms can be leveraged to actively control the spectral and temporal dynamics of QEs in the presence of graphene by modulating the free charge carrier concentration, e.g., through electrostatic gating~\cite{tielrooij2015electrical,cano2020fast,eriksen2022optoelectronic}. Complementing its active electrical tunability, the patterning of graphene into nanostructures supporting localized plasmon resonances enables further passive control over the associated near-fields~\cite{brar2013highly}. The coupling of QEs with localized graphene plasmons in one-dimensional graphene nanoribbons or zero-dimensional nanodisks is predicted to enable strong vacuum Rabi splittings \cite{koppens2011graphene}, superradiance~\cite{huidobro2012superradiance}, and efficient energy transfer~\cite{karanikolas2016tunable}.

In addition to the electrical tunability achieved by modulating charge carrier doping levels, a static magnetic field applied normally to a graphene sheet can further tune its optical response \cite{peres2006electronic,castroneto2009electronic}. In particular, the symmetry-breaking induced by a finite magnetic field induces a gyrotropic optical response in graphene~\cite{ferreira2011faraday,sounas2011electromagnetic}, leading to a impressively large Faraday rotation associated with light traversing the 2D layer~\cite{crassee2011giant}. Naturally, the applied magnetic field can also be leveraged as an additional knob to control the properties of graphene magnetoplasmon resonances operating at IR and THz frequencies~\cite{yan2012infrared}, which are anticipated to strongly enhance Faraday rotation~\cite{tymchenko2013faraday}, control spontaneous emission from a nearby QE~\cite{kortkamp2015active,ma2022local}, and drive a large chiroptical response~\cite{eriksen2025chiral}. However, the light emission characteristics of QEs interfacing localized magnetoplasmons in low-dimensional graphene nanostructures, including graphene nanoribbons and nanodisks, remains unexplored.

In this paper, we explore possibilities to actively control photon emission from QEs in the presence of extended or nanopatterned magneto-optical graphene monolayers. To do so, we develop rigorous semianalytical models to describe the quantum electrodynamic response of gyrotropic 2D materials and their low-dimensional forms, adopting the magneto-optical conductivity of graphene in the local limit of the random-phase approximation (RPA) to account for intra- and inter-band electronic transitions among the quantized Landau levels (LLs) induced by an applied static magnetic field~\cite{ferreira2011faraday}. Intriguingly, the induced LLs perturb the spectrally flat interband optical response of graphene at photon energies $\hbar\omega>2\EF$, producing sharp spectral features known as Shubnikov-de-Haas (SdH) oscillations that are associated with the quantization of cyclotron orbits~\cite{tan2011shubnikov}. Starting from the case of an extended graphene sheet in the presence of a perpendicularly applied static magnetic field, we study the effects of the induced Hall conductivity, magnetoplasmon polaritons, and SdH oscillations on the quantum electrodynamic response of a proximal QE. We then turn to low-dimensional structures by considering graphene nanoribbons supporting magnetoplasmon polaritons that are localized in one spatial dimension. Here, the interplay of geometrical near-field confinement and the intrinsic gyrotropic response of the 2D layer is found to produce large dissymmetry in the spontaneous decay of a left-hand circularly polarized (LCP) dipole versus that of a right-hand circularly polarized (RCP) dipole, indicating a large chiral response associated with the excitation of guided magnetoplasmon polaritons. This chiral dissymmetry is further enhanced in zero-dimensional nanodisks, leading to an even stronger distinction of LCP and RCP photon generation. Our findings hold promise for developing quantum light sources with actively controllable chiral degrees of freedom.

\section{Macroscopic quantum electrodynamics}

Our starting point is the general description of the quantum electrodynamic response of a QE in a photonic environment. More specifically, we consider a $N$-level QE, characterized by stationary states $\ket{j}$ with energies $\hbar\vep_j$, that is positioned at $\rb=(x,y,z)$ and interacting with the vacuum electromagnetic reservoir. In the macroscopic quantum electrodynamic formalism, the Hamiltonian governing this system is expressed as
\begin{align} \label{eq:H}
    \mathcal{H} = \hbar \sum_{j=1}^N\vep_j\ket{j}\bra{j} + \hbar \int_0^\infty{\rm d}\ww\,\ww \int{\rm d}^3\rb\,\hat{\bf f}_\ww^\dagger(\rb)\cdot\hat{\bf f}_\ww(\rb) \nonumber \\
    -\hat{\pb}\cdot\int_0^\infty{\rm d}\ww\,\sqpar{\hat{\Eb}_\ww(\rb) + \hat{\Eb}_\ww^\dagger(\rb)} ,
\end{align}
where $\hat{\bf f}_\ww^{\dagger}(\rb)$ and $\hat{\bf f}_\ww(\rb)$ are creation and annihilation operators, respectively, of electromagnetic excitations in the reservoir dressed by the photonic environment, $\hat{\pb}$ is the dipole operator, and
\begin{align}
    \hat{\Eb}_{\ww}(\rb) = \ii\mu_0 \ww \sqrt{\frac{\hbar\ww}{\pi}}&\int{\rm d}^3\rb' \Gm_\ww(\rb,\rb')  \nonumber \\
    \cdot&\int{\rm d}^3\rb''{\mathcal K}_\ww(\rb',\rb'')\cdot\hat{\bf f}_\ww(\rb'')
\end{align}
is the quantized radiation field operator, expressed in terms of the dyadic Green's tensor $\Gm_\ww$ and the generalized square root of the real part of the (three-dimensional) conductivity tensor $\mathfrak{Re}\{\sigma_\ww^{\rm 3D}(\rb,\rb')\}=\int{\rm d}^3\rb''{\mathcal K}_\ww(\rb,\rb'')\cdot {\mathcal K}_\ww^\dagger (\rb',\rb'')$~\cite{scheel2008macroscopic,buhmann2012macroscopic,buhmann2013dispersion,ge2013accessing,forati2014graphene, antao2021two,eriksen2022optoelectronic}. In particular, we define the generalized real and imaginary parts of a tensor $\Am(\rb,\rb')$ as $\mathfrak{Re}\clpar{\Am(\rb,\rb')}=\sqpar{\Am(\rb,\rb')+\Am^\dagger(\rb',\rb)}/2$ and $\mathfrak{Im}\clpar{\Am(\rb,\rb')}=\sqpar{\Am(\rb,\rb')-\Am^\dagger(\rb',\rb)}/2\ii$, respectively, such that they reduce to the usual real and imaginary parts when $\Am$ obeys Onsager reciprocity, i.e., $\Am(\rb,\rb')=\Am^{\rm T}(\rb',\rb)$, with the superscript ${\rm T}$ denoting transpose.

Following the standard procedure of tracing over the reservoir degrees of freedom and invoking the Born-Markov approximation, the master equation for the density matrix $\hat{\rho}$ describing electron dynamics in a two-level atom is obtained from the Hamiltonian of Eq.~\eqref{eq:H} as
\begin{equation}
    \pd{\hat{\rho}}{t} = \Gamma_{21}\rho_{22}\ket{1}\bra{1} - \sqpar{\ccpar{\frac{\Gamma_{21}}{2}+\ii\delta\vep_{21}}\ket{2}\bra{2}\hat{\rho} + {\rm H.c.}} ,
\end{equation}
where $\rho_{ij}\equiv\bra{i}\hat{\rho}\ket{j}$ denotes density matrix elements, from which the spontaneous emission rate
\begin{equation} \label{eq:SE}
    \Gamma_{ij} = \frac{2\mu_0}{\hbar} \vep_{ij}^2 \Imm\clpar{\pb_{ij}^* \cdot \Gm_{\vep_{ij}}(\rb, \rb) \cdot \pb_{ij}}
\end{equation}
and the Lamb shift
\begin{equation} \label{eq:Lamb_PV}
    \delta\vep_{ij} = \frac{\mu_0}{\pi\hbar}\mathcal{P} \int d\ww\frac{\ww^2}{\vep_{ij}-\ww}\Imm\clpar{\pb_{ij}^*\cdot\Gm_\ww(\rb,\rb) \cdot \pb_{ij}}
\end{equation}
corresponding to an atomic transition with frequency $\vep_{ij}\equiv\vep_i-\vep_j$ and dipole moment $\pb_{ij}\equiv\bra{i}\hat{\pb}\ket{j}$ naturally emerge, with $\mathcal{P}$ denoting the Cauchy principal value of the integral. Importantly, the $\Imm\{...\}$ appearing in Eqs.~\eqref{eq:SE} and $\eqref{eq:Lamb_PV}$ indicates the ``usual'' imaginary part of the scalar argument, and arises by noting that
\begin{equation}
    \pb_{ij}^* \cdot \mathfrak{Im}\clpar{ \Gm_{\vep_{ij}}(\rb, \rb) }\cdot \pb_{ij}=\Imm\clpar{\pb_{ij}^* \cdot \Gm_{\vep_{ij}}(\rb, \rb) \cdot \pb_{ij}} ,
\end{equation}
The distinction above is particularly important when considering the quantum electrodynamic response of circularly polarized dipoles.

Isolating the bare $\Gm_\ww^0$ and reflected $\Gm_\ww^{\rm ref}$ parts of the Green's dyadic $\Gm_\ww=\Gm_\ww^0+\Gm_\ww^{\rm ref}$, the spontaneous emission rate becomes
\begin{gather} \label{eq:Purcell}
    \Gamma_{ij} = \Gamma_{ij}^0 + \frac{2\mu_0}{\hbar} \vep_{ij}^2 \Imm\clpar{\pb_{ij}^* \cdot \Gm_{\vep_{ij}}^{\rm ref}(\rb, \rb) \cdot \pb_{ij}} ,
\end{gather}
where $\Gamma_{ij}^0 = \vep_{ij}^3\abs{\pb_{ij}}^2/3\pi\eps_0\hbar c^3$ is the vacuum emission rate \cite{novotny2012principles}. Using the same decomposition of the Green's dyadic, the Lamb shift can be reduced to
\begin{gather} \label{eq:Lamb}
    \delta\vep_{ij} \approx -\frac{\mu_0}{\hbar}\vep_{ij}^2 \Ree\clpar{\pb_{ij}^*\cdot\Gm^{\rm ref}_{\vep_{ij}}(\rb, \rb)\cdot\pb_{ij}}
\end{gather}
by absorbing the diverging contribution associated with $\Gm_\ww^0$ in the atomic transition frequency and approximating the Cauchy principle value integral by the real part of $\Gm_\ww^{\rm ref}$ \cite{dzsotjan2011dipole,eriksen2022optoelectronic}. The spontaneous emission rate and the Lamb shift thus characterize the quantum electrodynamic response of an arbitrary nanophotonic environment.

In what follows, we apply the above formalism to quantify the spontaneous emission rate $\Gamma$ and Lamb shift $\delta\vep$ associated with a two-level QE near extended or patterned graphene in the presence of a static perpendicular magnetic field $B$, as illustrated schematically in Fig.~\ref{fig1}(a,b) for a two-level atomic system positioned a distance $z$ above an extended graphene sheet. In the case of graphene doped to a Fermi energy $\EF$, the applied magnetic field drives cyclotron motion of charge carriers, leading to gyrotropic components in the 2D conductivity tensor and a transformation of the continuous low-energy Dirac cone electronic structure into a series of discrete Landau levels (LLs), as depicted in Fig.~\ref{fig1}(c). Importantly, the effect of the gyrotropic photonic environment on the quantum electrodynamic response is fully contained in the reflected part of the Green's dyadic, which we describe below for extended 2D materials, one-dimensional (1D) atomically thin waveguides, and zero-dimensional (0D) nanodisks.

\section{Magneto-optical response of extended graphene}

As depicted in Fig.~\ref{fig1}(a), a static magnetic field of magnitude $B$ applied perpendicularly to a graphene sheet induces Landau quantization of charge carriers undergoing cyclotron motion, leading to finite Hall conductivity. Specifically, the eigenenergies of carriers are quantized according to
\begin{equation} \label{eq:LL_energylevels}
    E_\ell = {\rm sign}(\ell)\frac{\hbar\vF}{L_B}\sqrt{2\abs{\ell}}
\end{equation}
for $\ell\in\{0,\,\pm1,\,\pm2,\,\dots\}$, where $\vF\approx c/300$ is the Fermi velocity in graphene and $L_B=\sqrt{\hbar/(eB)}$ is the magnetic length. Fig.~\ref{fig1}(c) depicts the energies of Landau levels overlaying the conical electronic dispersion of graphene at zero temperature doped to a Fermi energy $\EF$, with $\NF$ denoting the highest occupied level. As reported in Ref.~\cite{ferreira2011faraday}, the 2D magneto-optical conductivity of graphene is described in the local limit of the RPA as a sum over Landau level (LL) transitions with energies $E_{\ell\ell'}=E_\ell-E_{\ell'}$ according to
\begin{equation} \label{eq:sigmas}
    \sigma_{\parallel,\perp}(\ww) = \frac{\ii e^2}{h}\sum_{\substack{\ell,\ell'=-N_c\\\ell\neq \ell'}}^{N_c}\frac{\Lambda_{\ell\ell'}^{\parallel,\perp}}{E_{\ell\ell'}}\frac{f_{\ell'}-f_j}{\hbar\omega+E_{\ell\ell'}+\ii\tau^{-1}} ,
\end{equation}
where $\Lambda_{\ell\ell'}^\parallel = (\hbar\vF/L_B)^2(1+\delta_{\ell,0}+\delta_{\ell',0})\delta_{|\ell'|-|\ell|,\pm1}$, $\Lambda_{\ell\ell'}^\perp = \ii\Lambda_{\ell\ell'}^\parallel(\delta_{|\ell'|,|\ell|-1}-\delta_{|\ell'|-1,|\ell|})$, $\tau$ is a phenomenological lifetime associated with inelastic scattering, and $f_\ell$ denotes the occupation factor of the $\ell^{\rm th}$ LL determined from Fermi-Dirac statistics.

In the absence of a magnetic field, the doping charge carrier density $n$ in a graphene sheet at zero temperature is related to the Fermi energy through $\EF(n)=\hbar\vF\sqrt{\pi n}$. Thus, the occupation factors $f_\ell$ in the presence of a finite magnetic field can be determined by conserving the charge carrier number $n$ in graphene when $B=0$, which amounts to sequentially filling LLs---each with degeneracy $1/(2\pi L_B^2)$ per graphene area---until the condition
\begin{equation} \label{eq:carrier_number}
    \sum_{\ell=0}^{N_{\rm c}} (f_\ell-f_{-\ell}+1) = \frac{E_{\rm F}^2}{2\hbar\vF^2 eB}
\end{equation}
is satisfied. Incidentally, for finite temperature $T$, one may instead resort to solving Eq.~\eqref{eq:carrier_number} for the chemical potential $\mu$ entering the Fermi-Dirac occupation factors $f_\ell=\left[\ee^{(E_\ell-\mu)/\kB T}+1\right]^{-1}$, as described in Ref.~\cite{eriksen2025chiral}. Alternatively, the occupation factors can be determined by specifying the Fermi energy $\EF$ for $B\neq0$, implying that the considered graphene system interfaces a ``reservoir'' of charge carriers. In what follows, we contrast results obtained in either of these configurations, i.e., by specifying $\EF(n)$ for the carrier density $n$ corresponding to $B=0$, or assigning $\EF$ when $B\neq 0$. In practice, the summations over Landau levels in Eqs.~\eqref{eq:sigmas} and \eqref{eq:carrier_number} are performed up to $N_{\rm c}={\rm int}\{(E_{\rm c}/E_1)^2\}$ for a cutoff energy $E_{\rm c} = 2.7$\,eV, i.e., approximating the nearest-neighbor tight-binding hopping energy of graphene, which leads to well-converged results for the relevant parameters considered here \cite{ferreira2011faraday}.

\section{Light emission mediated by a two-dimensional magneto-optical graphene sheet}

Invoking the electromagnetic boundary conditions arising from Maxwell's equations, we extend the approach of Refs.~\cite{hohenester2020nano,eriksen2022optoelectronic} to determine the reflected part of the Green's tensor. Specifically, we consider the self-interaction of a point dipole at $\rb$ in a homogeneous isotropic medium above a planar interface characterized by the reflection coefficients $r_{\alpha\beta}$ associated with incoming (outgoing) $\alpha$-polarized ($\beta$-polarized) light for $\alpha,\beta\in \{{\rm s},{\rm p}\}$, thus accounting for cross-polarization of transverse electric (TE) or ``senkrecht'' s-polarized and transverse magnetic (TM) or ``parallel'' p-polarized electromagnetic fields interacting with a gyrotropic medium. For a gyrotropic 2D material characterized by local longitudinal $\sigma_\parallel$ and Hall $\sigma_\perp$ conductivities that interface isotropic dielectric media with permittivity (permeability) $\eps_1$ ($\mu_1$) above and $\eps_2$ ($\mu_2$) below, the introduction of a 2D current in the electromagnetic boundary conditions leads to the Fresnel reflection coefficients~\cite{kortkamp2015active}
\begin{subequations}
\begin{align}\label{eq:Fresnel_coeff_single_general_interface}
    r_{\rm ss} &= \frac{\Dm_{\eps}^{(+)} \Dm_{\mu}^{(-)} - \sigma_\perp^2}{\Dm_{\eps}^{(+)} \Dm_{\mu}^{(+)} + \sigma_\perp^2} ,  \\
    r_{\rm pp} &= -\frac{\Dm_{\eps}^{(-)} \Dm_{\mu}^{(+)} - \sigma_\perp^2}{\Dm_{\eps}^{(+)} \Dm_{\mu}^{(+)} + \sigma_{\perp}^2} ,  \\
    r_{\rm sp} &= r_{\rm ps} = -\frac{2}{Z_1} \frac{\sigma_\perp}{\Dm_{\eps}^{(+)} \Dm_{\mu}^{(+)} + \sigma_\perp^2} ,  
\end{align}
\end{subequations}
where $Z_{\rm m}=\sqrt{\mu_{\rm m}/\eps_{\rm m}}$ is the impedance in medium ${\rm m}$, and we have defined
\begin{subequations}
\begin{align}
    \Dm_{\eps}^{(\pm)} &\equiv \frac{\eps_1}{k_{1z}} \pm \frac{\eps_2}{k_{2z}} \pm \frac{\sigma_\parallel}{\ww} ,  \\
    \Dm_{\mu}^{(\pm)} &\equiv \frac{k_{1z}}{\mu_1} \pm \frac{k_{2z}}{\mu_2} \pm \ww \sigma_\parallel .
\end{align}
\end{subequations}
In Appendix~\ref{app:Fresnel} we present a succinct derivation of the above Fresnel reflection and transmission coefficients for arbitrary plane waves impinging on a gyrotropic 2D material.

Using the reflection coefficients in Eqs.~\eqref{eq:Fresnel_coeff_single_general_interface} above, we decompose the electromagnetic fields in the conserved in-plane wave vector components $k_\parallel$ according to Appendix~\ref{app:Greens} to obtain the Green's tensor
\begin{widetext}
\begin{equation} \label{eq:G_ref_gyrotropic}
    \Gm_\ww^{\rm ref} = \frac{\ii}{8\pi} \int_{0}^{\infty}{\rm d}k_{\parallel} k_{\parallel} \frac{\ee^{2\ii k_{1z}z}}{k_1^2k_{1z}}  \begin{bmatrix} k_1^2r_{\rm ss}-k_{1z}^2r_{\rm pp} & -k_1 k_{1z} (r_{\rm sp}+r_{\rm ps}) & 0  \\
    k_1 k_{1z} (r_{\rm sp}+r_{\rm ps}) & k_1^2r_{\rm ss}-k_{1z}^2r_{\rm pp} & 0  \\
    0 & 0 & 2k_{\parallel}^2 r_{\rm pp}
    \end{bmatrix} ,
\end{equation}
\end{widetext}
where $k_{{\rm m}z} = \sqrt{\mu_{\rm m}\eps_{\rm m}\ww^2-k_\parallel^2+\ii 0^+}$ denotes the normal component of the wave vector $\kb_{\rm m}$ in a medium with permittivity (permeability) $\eps_{\rm m}$ ($\mu_{\rm m}$). The integration in Eq.~\eqref{eq:G_ref_gyrotropic} is evaluated following the procedure in Ref.~\cite{paulus2000accurate}.

\begin{figure*}
    \centering
    \includegraphics[width=\textwidth]{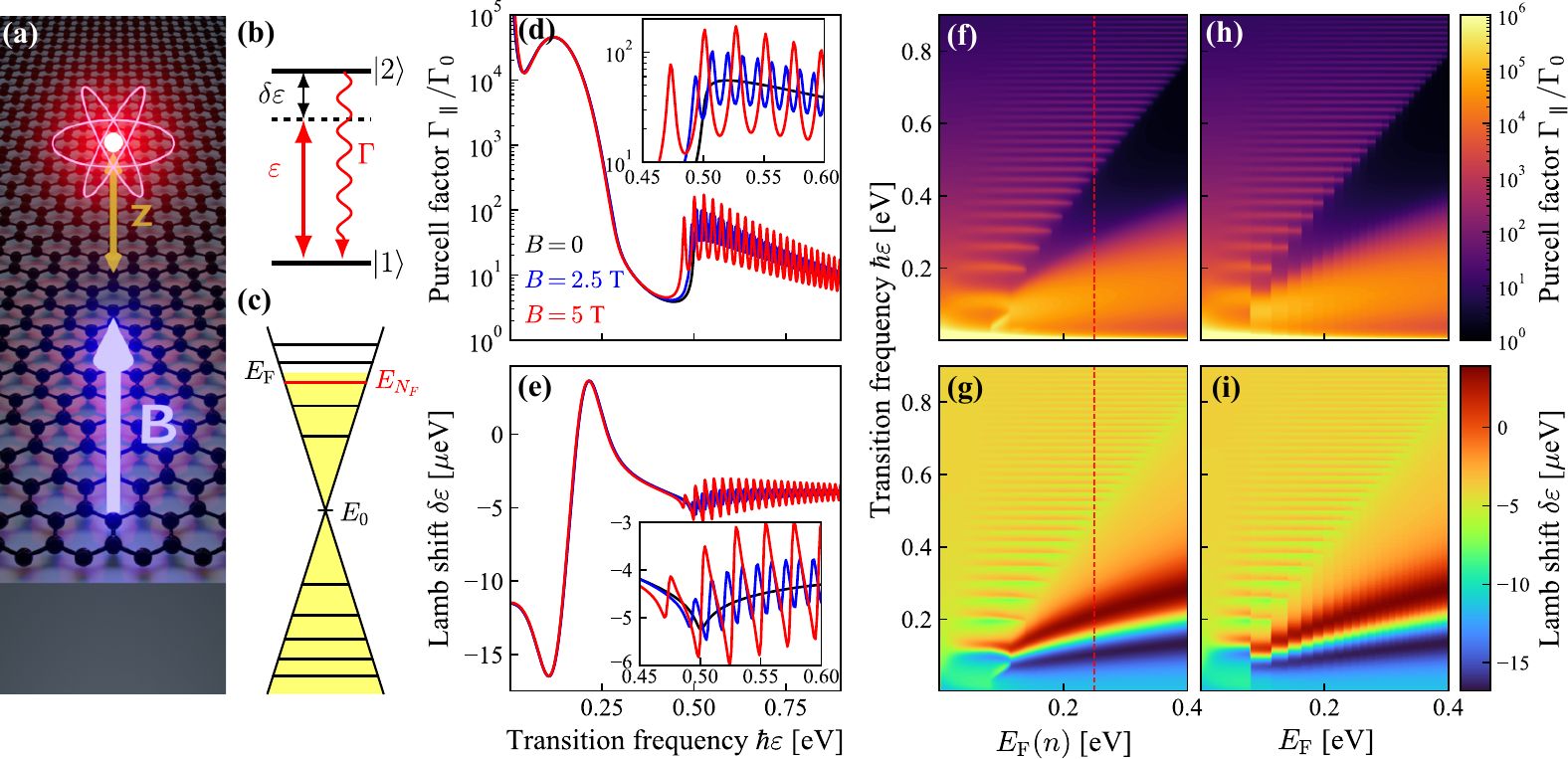}
    \caption{{\bf Electrical manipulation of a quantum light emitter near an extended magneto-optical graphene sheet.} (a) Illustration of a quantum emitter (QE) located a distance $z$ above a graphene sheet on a dielectric substrate that is subjected to an external magnetic field of amplitude $B$ oriented normally to the graphene plane. (b) Energy level diagram of a two-level atom with bare transition frequency $\vep$, exhibiting a Lamb shift $\delta\vep$ and enhanced spontaneous emission at the rate $\Gamma$ in the presence of graphene. (c) Schematic of the conical graphene electron dispersion relation (dashed lines), populated up to Fermi energy $\EF$ for $B=0$, that is fragmented in discrete Landau levels (horizontal solid lines), populated up to level $N_{\rm F}$ with energy $E_{N_{\rm F}}$, in the presence of a static magnetic field. The Purcell factor (d) and Lamb shift (e) corresponding to a dipole with transition dipole moment $p=1\,e\cdot$nm at $z=25$\,nm above and oriented parallel to a graphene sheet with fixed carrier density $n\approx 4.6 \times 10^{12}$\,cm$^{-2}$ (Fermi energy $\EF=0.25$\,eV for $B=0$) are plotted for $B=0$ (black curves), $B=2.5$\,T (blue curves), and $B=5$\,T (red curves) as a function of the dipole transition energy $\hbar\vep$. The inset shows the Purcell factors around the graphene interband transition threshold $\hbar\vep\sim 2\EF$. Under the same conditions, spectra of the Purcell factor (f) and Lamb shift (g) at $B=5$\,T are shown as functions of the $B=0$ Fermi energy, i.e., by fixing the carrier density, with the vertical dashed lines indicating the corresponding results shown in (d) and (e). The results in panels (h) and (i) are obtained by repeating the calculations in (f) and (g) by specifying the Fermi energy in the magneto-optical graphene sheet. All results are obtained by setting the phenomenological inelastic scattering rate in graphene to $\tau^{-1}=e v_{\rm F}^2/\mu_{\rm DC}\EF$ with DC mobility $\mu_{\rm DC}=10^4$\,cm$^2/$Vs, choosing the substrate permittivity $\eps_2=2\eps_0$, and assuming zero temperature.}
    \label{fig1}
\end{figure*}

The combined formalism above describes the quantum electrodynamic response of a magneto-optical graphene sheet. We apply this formalism to simulate the light emission characteristics of a nearby QE, illustrated schematically in Fig.~\ref{fig1}(a) for a graphene sheet in the $x$-$y$ plane and subjected to a perpendicular static magnetic field. Specifically, for a transition dipole oriented parallel to the graphene sheet and located a distance $z=25$\,nm above it, we present simulations of the Purcell factor quantifying the QE emission rate and the Lamb shift of the QE transition energy in Figs.~\ref{fig1}(d) and \ref{fig1}(e), respectively, for zero magnetic field (black curves) and finite $B=2.5$\,T (blue curves) or $B=5$\,T (red curves) fields. For the $\EF=0.25$\,eV charge carrier doping level considered, both the Purcell factor and Lamb shift display prominent low-energy features that are independent of the applied magnetic field and attributed to plasmon resonances that strongly enhance near-field interactions. At higher energies $\hbar\vep>2\EF$, the continuum of interband transitions between occupied and unoccupied states for $B=0$ is instead comprised of sharp SdH oscillations associated with transitions between discrete Landau levels produced by the applied magnetic field.

In tandem with the applied magnetic field, the optical response of graphene can be actively tuned by varying the doping charge density. However, for a graphene sheet with a fixed charge carrier density $n$, the Fermi energy (or chemical potential at finite temperature) must adapt to changes in the electronic structure induced by the magnetic field. In Fig.~\ref{fig1}(f,g), we present simulations of the quantum electrodynamic response as a function of the Fermi energy while conserving the charge carrier density according to Eq.~\eqref{eq:carrier_number}. At low energies, prominent plasmonic features vary smoothly with $\sqrt{\EF}$, while interband transitions between LLs are quenched due to Pauli blocking above the $\hbar\vep>2\EF$ threshold. In an alternative scenario, we may consider a situation where the Fermi energy is fixed, such as in electrostatically gated graphene that has access to a reservoir of charge carriers, in which case the density can adjust according to the applied magnetic field. Fig.~\ref{fig1}(h,i) showcases the Purcell factor and Lamb shift as a function of Fermi energy with flexible carrier density, in which case the spectra exhibit abrupt jumps in both the plasmon resonance and in the SdH oscillations. This behavior is reminiscent of the spontaneous emission reported in Ref.~\cite{kortkamp2015active}, which also theoretically explores the interaction of a QE close to an extended magnetooptical graphene sheet. Notably, the results reported in Ref.~\cite{kortkamp2015active} are obtained by specifying the Fermi energy while varying the magnetic field, implying that the charge carrier density in graphene is not conserved, and resulting in the rapid jumps in the Purcell factor around $\EF\approx E_\ell$ that we also observe under fixed Fermi energy. Incidentally, these jumps become sharper at low temperatures, eventually leading to a discontinuous response at zero temperature when the occupation $f_\ell$ of LL $\ell$ jumps from 0 to 1 at $\EF=E_\ell$. Thus, at low temperatures, a magneto-optical graphene sheet could be used to detect slight variations in the magnetic field strength or Fermi energy.

In the presence of an external magnetic field, the nonvanishing off-diagonal Green's tensor elements in Eq.~\eqref{eq:G_ref_gyrotropic} suggest a distinguishable response to LCP and RCP dipoles $\pb_{\rm LCP}=p(\xx+\ii \yy)/\sqrt{2}$ and $\pb_{\rm RCP}=p(\xx-\ii \yy)/\sqrt{2}$, respectively, i.e., of equal magnitude $p$, potentially enabling tunable chiral light emission of QEs near a magneto-optical graphene sheet. In the spirit of circular dichroism under plane wave excitation, we characterize chiral light-matter interactions in the graphene-QE system by simulating the decay rates of LCP versus RCP dipoles. A natural figure of merit that quantifies this difference is the dissymmetry factor
\begin{gather} \label{eq:dissymmetryfactor_g_definition}
    g=2\frac{\Gamma_{\rm LCP}-\Gamma_{\rm RCP}}{\Gamma_{\rm LCP}+\Gamma_{\rm RCP}} ,
\end{gather}
where $\Gamma_{\alpha}\propto \Imm\{\pb_{\alpha}^*\cdot\Gm_\vep^{\rm ref}\cdot\pb_\alpha\}$ is the decay rate of a $\alpha\in\{{\rm LCP},{\rm RCP}\}$ dipole. The dissymmetry factor is per definition restricted to $-2\leq g\leq 2$. However, for the parameters considered in Fig.~\ref{fig1}(d), we find small dissymmetry factors restricted to below $|g|\leq 0.001$, thus motivating explorations of nanostructured materials that produce large spatial variations in the electromagnetic near-field on commensurate length scales to obtain a larger chiral near-field response.

\begin{figure}
    \centering
    \includegraphics[width=\linewidth]{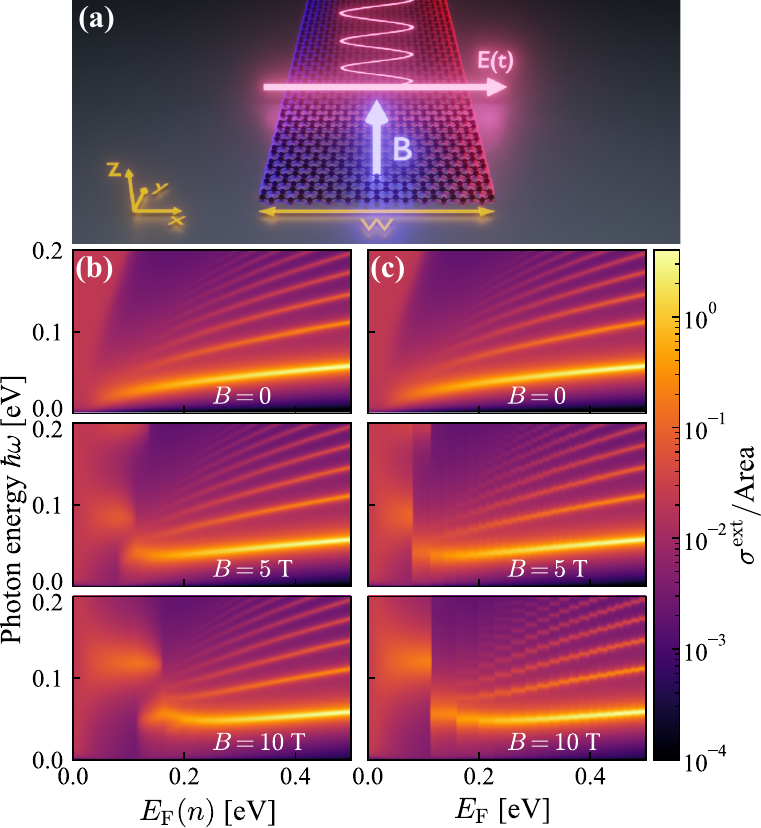}
    \caption{{\bf Far field response of localized magnetoplasmons in graphene ribbons.} (a) Illustration of magnetoplasmons in a graphene nanoribbon excited by a plane wave  impinging normally to the graphene plane with electric field linearly polarized along $\hat{\xb}$. Extinction cross sections of a $W=1$\,$\mu$m graphene ribbon are shown for the magnetic field strengths indicated in each panel as a function of (b) charge carrier density specified by the graphene Fermi energy for $B=0$ and (c) fixing the Fermi energy when $B\neq0$. Here the phenomenological damping rate $\tau^{-1}=e v_{\rm F}^2/\mu_{\rm DC} \EF$ is specified for a DC mobility of $\mu_{\rm DC}=5000$\,cm$^2/$Vs, the substrate permittivity is $\eps_2=\eps_0$, and temperature effects are neglected.}
    \label{fig2}
\end{figure}

\section{One-dimensional magneto-optical graphene nanoribbons}

We consider a 2D ribbon with finite width $W$ in the $x$-direction, infinite extension in $\yy$, and an intrinsic gyrotropic conductivity. To describe light-matter interactions in this system, we generalize the formalism reported in Refs.~\cite{christensen2015kerr,rasmussen2023nonlocal}, based on self-consistent solution of the scalar potential and applied to graphene nanoribbons exhibiting an isotropic intrinsic optical response, to obtain the self-consistent current density that accounts for the tensorial nature of the gyrotropic conductivity.

\begin{figure*}
    \centering
    \includegraphics[width=\linewidth]{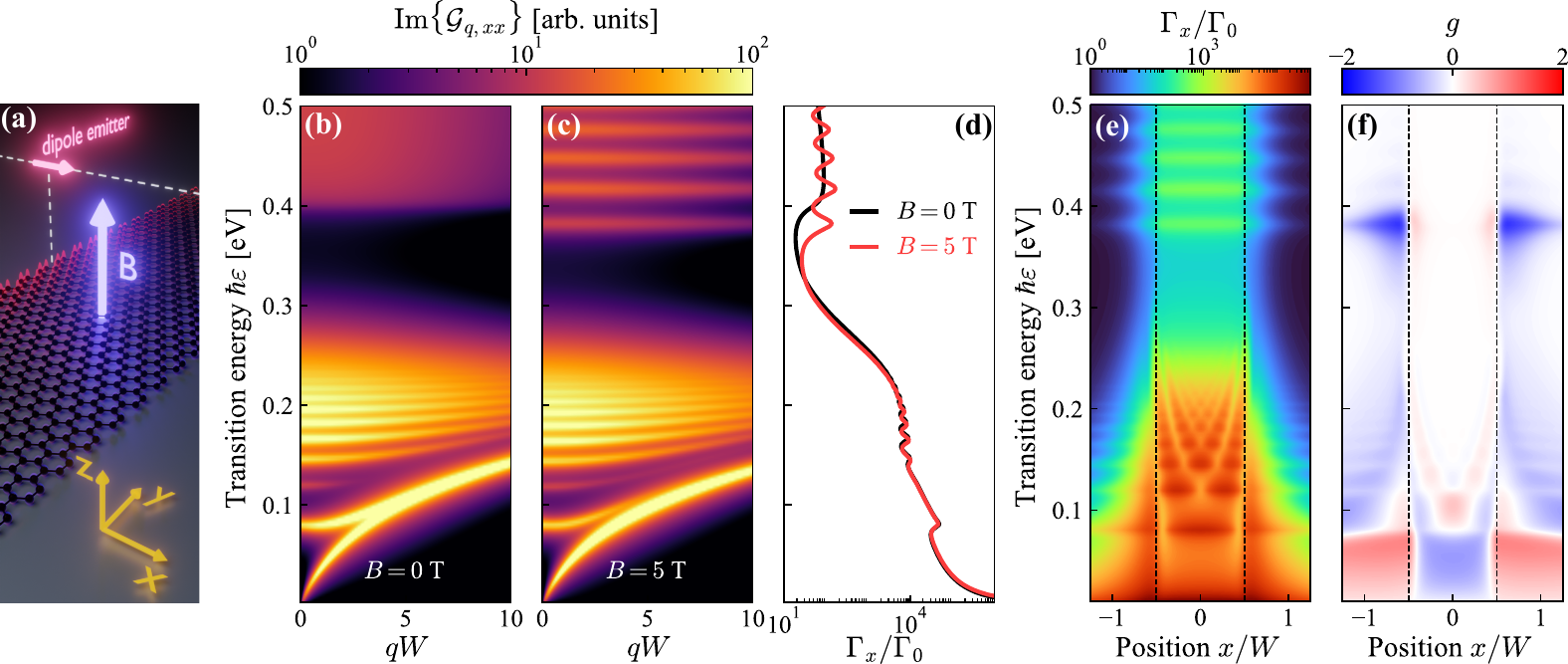}
    \caption{{\bf Dipolar excitation and chiral near-field response of magneto-optical graphene ribbons.} (a) Illustration of a graphene ribbon in the $z=0$ plane subjected to a perpendicular static magnetic field of magnitude $B$ and interacting with a nearby point dipole emitter. The wave-vector-resolved Green's tensor, $\Gm_{q,xx}$, is plotted for (b) $B=0$ and (c) $B=5$\,T as a function of the wave vector $q$ along the direction of translational invariance and emitter transition energy, while the corresponding Purcell factors (obtained after integrating $(2\pi)^{-1}\int{\rm d q}\Gm_{q,xx}$) are presented in panel (d). Note that the contour plots in panels (b) and (c) are saturated to highlight the plasmon dispersion, and in all cases the dipole is oriented along $\xx$ and located 25\,nm directly above the ribbon edge. The spectral dependence of (e) the Purcell factor $\Gamma_x/\Gamma_0$ for a dipole polarized in $\xx$ and (f) the chiral dissymmetry factor $g$ given by Eq.~\eqref{eq:dissymmetryfactor_g_definition} for circularly polarized dipoles are shown as a function of the dipole position along $x$ while maintaining a height $z=25$\,nm above the ribbon when it is subjected to a magnetic field of $B=5$\,T. All results here are obtained for a ribbon of width $W=200$\,nm doped with a charge carrier density corresponding to $\EF=0.2$\,eV for $B=0$ and mobility $5000$\,cm$^2/$Vs.}
    \label{fig3}
\end{figure*}

In the framework of the quasistatic approximation, we write the total time-harmonic electric field oscillating with frequency $\omega$ acting on the $\Rb=(x,y)$ plane within a 2D material as
\begin{equation} \label{eq:E_E_rho}
    \Eb(\Rb) = \Eb^{\rm ext}(\Rb) - \frac{1}{4\pi\epseff}\nabla_\Rb\int {\rm d}^2\Rb'\frac{\rhoind(\Rb')}{\abs{\Rb-\Rb'}} ,
\end{equation}
where the first and second terms account for external and induced fields, respectively, the latter expressed in terms of the induced charge density $\rhoind$ and the effective permittivity $\epseff=(\eps_1+\eps_2)/2$ obtained by averaging the permittivity above and below the ribbon. Inserting the field above into Ohm's law for the current $\Jb=\sigma\cdot\Eb$ and substituting the charge density for the current according to the continuity equation $\rhoind = (-\ii/\ww)\nabla_\Rb\cdot\Jb$, we obtain the self-consistent equation
\begin{align} \label{eq:J_J}
    \Jb(\Rb) &= \sigma(\Rb)\cdot\Eb^{\rm ext}(\Rb) \nonumber \\
    & + \frac{\ii\sigma(\Rb)}{4\pi\epseff\ww}\cdot\nabla_{\Rb}\otimes\nabla_{\Rb} \cdot\int{\rm d}^2\Rb'\frac{\Jb(\Rb')}{|\Rb-\Rb'|} ,
\end{align}
after integrating by parts and replacing $\nabla_{\Rb'}\left\vert\Rb-\Rb'\right\vert^{-1}\to-\nabla_{\Rb}\left\vert\Rb-\Rb'\right\vert^{-1}$ inside the integral. To solve for the current, we write the conductivity as $\sigma(\Rb)=f(x)\sigma(\ww)$, where $f(x)$ is a geometrical factor that is unity within the 2D material but zero elsewhere and $\sigma(\ww)$ is the (local) intrinsic conductivity tensor of the 2D material, and exploit the translational invariance in $\yy$ by decomposing the current density in plane waves $\Jb(\Rb)=(2\pi)^{-1}\int{\rm d}q\Jb_q(x)\ee^{\ii q y}$ with optical wave vector $q$. Multiplying the resulting expression by $\ee^{-\ii q y}$ and integrating over $y$, the Fourier components of the current density are obtained as
\begin{equation} \label{eq:J_E_DVJ}
     \Jb_q(x) = f(x)\sigma(\ww) \cdot \left[ \Eb^{\rm ext}_q(x) + \frac{\ii}{4\pi\epseff\ww}\Dm_q\Vm_q\Jb_q(x) \right] ,
\end{equation}
where $\Eb_q^{\rm ext}(x)=\int{\rm d}y\Eb^{\rm ext}(\Rb)\ee^{-\ii q y}$,
\begin{equation}
    \Dm_q\Jb_q(x) = \begin{pmatrix}
         \partial_x^2 & \ii q \partial_x \\ 
         \ii q \partial_x & -q^2
     \end{pmatrix} \cdot \Jb_q (x)
\end{equation}
contains the differentiation operators, and
\begin{equation}
    \Vm_q\Jb_q(x) = 2\int{\rm d}x' K_0\ccpar{|q||x-x'|}\Jb_q(x')
\end{equation}
is the integration operator, the latter obtained using the identity $\int_{-\infty}^\infty{\rm d}y'\left\vert\Rb-\Rb'\right\vert^{-1}\ee^{\ii q y'}=2K_0\ccpar{|q||x-x'|}\ee^{\ii q y}$ involving the modified Bessel function $K_0$ \cite{NIST:DLMF}. Following the real-space discretization approach outlined in Appendix~\ref{app:1D}, we can cast Eq.~\eqref{eq:J_E_DVJ} as a matrix equation to isolate the current
\begin{equation} \label{eq:J_final}
    \Jb_q(x) = \Mm_{q,\ww}(x) \cdot f(x)\sigma(\ww) \cdot\Eb_q^{\rm ext}(x) ,
\end{equation}
where we introduce the operator
\begin{equation}
    \Mm_{q,\ww}(x) = \left[\mathds{1}-\frac{\ii f(x) \sigma(\ww)}{4\pi\epseff\ww}\Dm_q\Vm_q\right]^{-1} .
\end{equation}
Incidentally, Eq.~\eqref{eq:J_E_DVJ} reduces to an eigenvalue equation in the absence of an external field, from which one can obtain the eigenmodes $\Jb_q(x)$ of the ribbon geometry.

Applying the formalism above, we first investigate the optical response of magneto-optical GNRs to free space illumination characterized by the external field $\Eb^{\rm ext}(\Rb)=E_0\xx$, i.e., a linearly-polarized plane wave with amplitude $E_0$, as illustrated schematically in Fig.~\ref{fig2}(a). More specifically, we insert the external field $\Eb^{\rm ext}_q=2\pi E_0\delta(q)\xx$ in Eq.~\eqref{eq:J_final} to compute the induced dipole moment $p_x=\int{\rm d}xx\rhoind(x)=(\ii/\ww)\int{\rm d}xJ_x(x)$, expressed in terms of the current through the continuity equation and integration by parts, from which we extract the polarizability $\alpha$ (from the definition $\pb=\alpha\cdot\Eb^{\rm ext}$) and the extinction cross section $\sigma^{\rm ext}=(\ww/c\sqrt{\eps_0\epseff})\Imm\{\alpha\}$~\cite{hohenester2020nano}. Here, the broken translational symmetry in $\xx$ associated with the nanoribbon geometry facilitates the excitation of localized graphene plasmon resonances that manifest as prominent features in the associated extinction cross section.

In Fig.~\ref{fig2}(b,c), we explore the tunability of a magneto-optical GNR of width $W=1$\,$\mu$m with Fermi energy, considering the cases of fixed and varying carrier density $n$ in Figs.~\ref{fig2}(b) and \ref{fig2}(c), respectively. In the absence of magnetic field, plasmons evolve monotonically with $\sqrt{\EF}$, while bright modes appearing in the spectra diminish with increasing mode order. In the presence of finite magnetic fields, the optical response in Fig.~\ref{fig2}(b) exhibits complex behavior around the interband region at low Fermi energies as each LL is continuously populated according to the imposed conservation of charge carriers. Similar behavior is seen in Fig.~\ref{fig2}(c), where results are shown by conserving the Fermi energy rather than carrier density, albeit with sharper changes in the considered zero-temperature limit when the Fermi energy intersects a LL, leading to dramatic changes of the allowed intraband transitions in the magneto-optical graphene nanoribbon.

To quantify the near-field of a magneto-optical graphene nanoribbon, we simulate its response to an external point dipole with moment $\pb$ located at $\rb_0$ that generates a field
\begin{equation} \label{eq:E_ext_dipole}
    \Eb^{\rm ext}(\Rb) = \frac{1}{4\pi\eps_1}\nabla_{\rb_0}\otimes\nabla_{\rb_0}\frac{1}{\abs{\Rb-\rb_0}}\cdot\pb
\end{equation}
at coordinates $\Rb$ in the ribbon. More specifically, as outlined in Appendix~\ref{app:1D}, the external field is decomposed in spatial Fourier components along $\yy$ to compute the corresponding components of the Green's dyadic
\begin{equation} \label{eq:G_q}
    \Gm_{q,\ww}^{\rm ref}(\rb_0,\rb_0) = \int{\rm d}x\tensor{g}_q^*(x,\rb_0)\cdot\Mm_{q,\ww}(x)\cdot\tensor{g}_q(x,\rb_0) ,
\end{equation}
where
\begin{equation} \label{eq:g_q}
    \tensor{g}_q(x,\rb_0) = 2\nabla_{\rb_0}\otimes\nabla_{\rb_0}K_0\sqpar{|q|\sqrt{(x-x_0)^2+z_0^2}}\ee^{-\ii q y_0} ,
\end{equation}
from which the full Green's tensor is constructed as $\Gm_\ww^{\rm ref}(\rb,\rb)\propto(2\pi)^{-1}\int{\rm d}q\Gm_{q,\ww}^{\rm ref}(\rb,\rb)\ee^{\ii q y}$ to determine the LDOS and the Purcell factor using Eq.~\eqref{eq:Purcell}. For a nanoribbon of $W=200$\,nm width doped with a fixed carrier density $n\approx 2.94 \times 10^{12}$\,cm$^{-2}$ ($\EF=0.2$\,eV in pristine graphene), the wave-vector-resolved Green's function corresponding to a dipole located at $\rb=(-100,0,25)$\,nm, i.e., 25\,nm directly above the edge of the ribbon, is plotted in Fig.~\ref{fig3}(b) and \ref{fig3}(c) for zero and $B=5$\,T perpendicular applied magnetic field strengths, respectively. In both cases, the dipole couples to numerous guided plasmon modes in the nanoribbon at energies below the $2\EF$ interband transition threshold. In analogy to the results presented in Fig.~\ref{fig1}(d,e), the static magnetic field produces small quantitative changes in the plasmon modes while inducing noticeable SdH oscillations in the interband response. These behaviors persist after integrating the Fourier components of the Green's function to obtain the Purcell factors shown in Fig.~\ref{fig3}(d).

\begin{figure*}
    \centering
    \includegraphics[width=\linewidth]{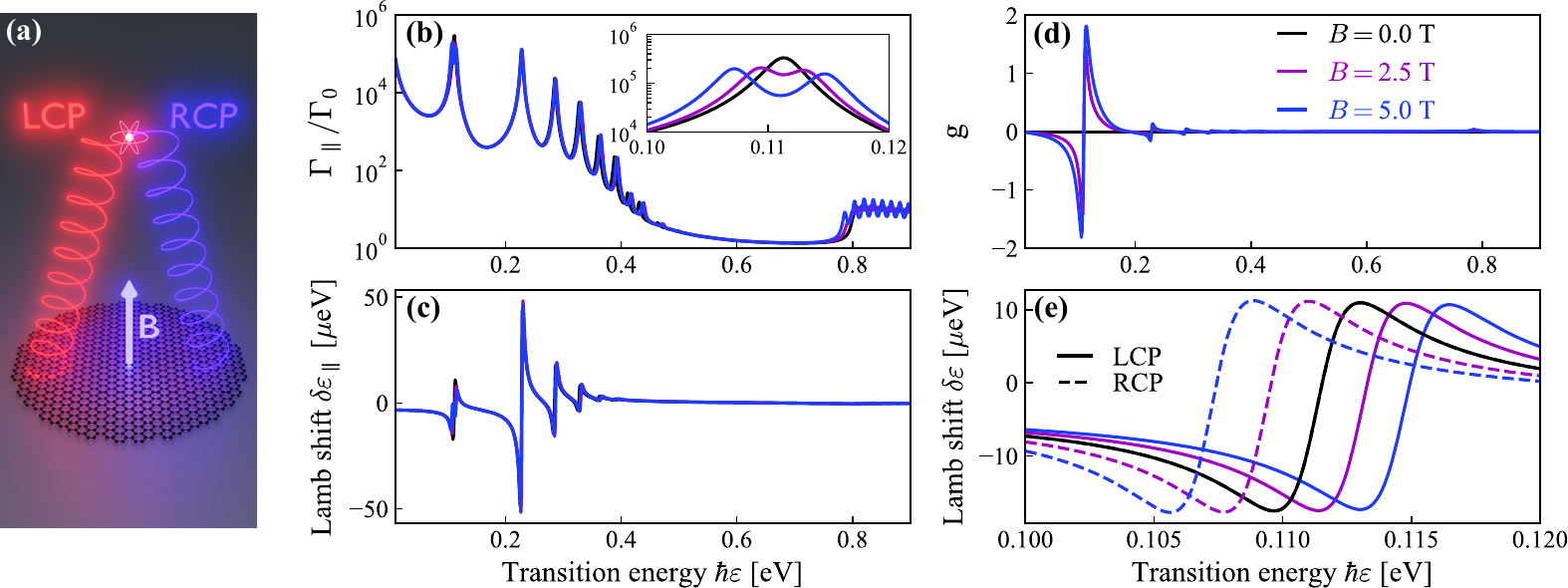}
    \caption{{\bf Active tuning of linearly and circularly polarized quantum light emitters mediated by a magneto-optical graphene disk.} (a) Illustration of a quantum emitter of left-circularly (LCP) and right-circularly (RCP) polarized photons located directly above a magnetooptical graphene disk. (b) Purcell factor and (c) Lamb shift corresponding to a dipole directly above and oriented parallel to a graphene disk at $B=0$ (black curve), $B=2.5$\,T (purple curve) and $B=5$\,T (blue curve). (d) The dissymmetry factor of Eq.~\eqref{eq:dissymmetryfactor_g_definition} quantifying chiral photon generation is plotted for the same parameters as in panels (b) and (c). (e) Lamb shift for a RCP (dashed curves) and a LCP (solid curves) dipole emitter with transitions frequencies near the dipolar plasmonic resonance. In all cases the dipole emitters are located a distance $z=30$\,nm directly above the center of a disk with diameter $D=200$\,nm, charge carrier doping density corresponding to $\EF=0.4$\,eV for $B=0$, mobility $\mu_{\rm DC}=5000$\,cm$^2/$Vs, and at zero temperature.}
    \label{fig4}
\end{figure*}

The spatial inhomogeneity of the localized fields associated with magnetoplasmon modes in graphene nanoribbons leads to a strong dependence of the light emitted by a dipole on its position. We explore this concept in Fig.~\ref{fig3}(e) by plotting the Purcell factor spectra obtained by sweeping the $x$-coordinate of a dipole across the ribbon at a fixed distance of $z=25$\,nm above the graphene plane. These results reveal the positions at which an emitter can most effectively launch guided plasmon modes, while features due to SdH oscillations at higher energies are relatively weaker and less sensitive to the dipole position. In Fig.~\ref{fig3}(f), a similar analysis is performed for circularly-polarized dipoles, where the difference in emission of left- and right-hand circularly-polarized light, quantified by the dissymmetry factor in Eq.~\eqref{eq:dissymmetryfactor_g_definition}, reveals the potential of magneto-optical graphene nanoribbons for controlling chiral light emission. Here, we achieve dissymmetry factors ranging from $g\approx-0.79$ to $g\approx 0.84$ when a QE is located immediately above the ribbon, while dissymmetry factors down to $g\approx -1.57$ emerge in the region \textit{outside} the ribbon for transition energies approaching the $\hbar\vep\sim 2\EF$ interband threshold and SdH oscillations. 

\section{Zero-dimensional magneto-optical graphene nanodisks}

To consider a further reduction in the dimensionality of the 2D gyrotropic material, we turn a 0D magneto-optical graphene nanodisk that supports magnetoplasmons localized in two spatial dimensions. To quantify the near-field response, we adapt the semianalytical formalism developed in Ref.~\cite{eriksen2025chiral} to compute the polarizability of a gyrotropic nanodisk but instead consider the interaction with a point dipole. Specifically, we express the scalar potential $\Phi$ acting on a dipole located at $\rb_0$ in a medium with permittivity $\eps_1$ as
\begin{equation} \label{eq:Phiind_rho}
    \Phi^{\rm ind}(\rb_0) = \frac{1}{4\pi\epseff}\int{\rm d}^2\Rb\frac{\rhoind(\Rb)}{\abs{\rb_0-\Rb}} ,
\end{equation}
where the induced charge density $\rhoind$ is evaluated at points $\Rb$ within the disk. For a disk with diameter $D$, it is convenient to move to cylindrical coordinates, defined by the radial distance $r$ and azimuthal angle $\varphi$, and expand the spatial dependence of the Coulomb interaction as
\begin{equation}\label{eq:Coulomb_int_expansion}
    \frac{1}{\abs{\rb-\rb'}} = \sum_{l=-\infty}^\infty\int_0^\infty{\rm d}k J_{|l|}(kr) J_{|l|}(kr')\ee^{\ii l(\varphi-\varphi')}\ee^{-k(z_>-z_<)} ,
\end{equation}
where $z_<={\rm min}(z,z')$ and $z_>={\rm max}(z,z')$. The induced charge density entering Eq.~\eqref{eq:Phiind_rho} is determined from Poisson's equation $\nabla^2\Phi=-\rhoind/\epseff$ for the total potential $\Phi(\Rb)=\Phi^{\rm ext}(\Rb)+\Phi^{\rm ind}(\Rb)$, expressed as a sum of external and induced contributions, with the latter term given by evaluating Eq.~\eqref{eq:Phiind_rho} at points within the disk. To obtain the 2D charge density, we expand it in Jacobi polynomials $P_j^{(|l|,0)}(\tilde{r})$ according to \cite{fetter1986magnetoplasmons,christensen2014classical}
\begin{equation} \label{eq:rho_l_Jacobi_expansion}
    \rho(\tilde{r},\varphi) = \sum_{l=-\infty}^{\infty}\sum_{j=0}^{\infty} a_j^{(l)} \tilde{r}^{|l|} P_j^{(|l|,0)} (1-2\tilde{r}^2) \ee^{\ii l\varphi} ,
\end{equation}
where $\tilde{r}=2r/D$ is the normalized radial distance and $a_j^{(l)}$ are expansion coefficients. By exploiting the orthonormal properties of Jacobi polynomials and their relations with Bessel functions, we construct a matrix equation for the vector ${\bf a}^{(l)}$ containing the expansion coefficients $a_j^{(l)}$ introduced above as
\begin{equation}
    \sqpar{\frac{\ii\epseff\ww D}{\sigma_\parallel}\ccpar{\frac{\sigma_\parallel-\ii{\rm sgn}(l)\sigma_\perp}{\sigma_\parallel+\ii{\rm sgn}(l)\sigma_\perp}\mathcal{A}^{(l)}+\mathcal{B}^{(l)}}+\mathcal{C}^{(l)}}{\bf a}^{(l)}={\bf b}^{(l)} ,
\end{equation}
where the matrices $\mathcal{A}^{(l)}$, $\mathcal{B}^{(l)}$, and $\mathcal{C}^{(l)}$, defined in Appendix~\ref{app:disk}, operate on the vector ${\bf b}^{(l)}$ containing entries
\begin{equation}
    b_j^{(l)} = \frac{4\epseff}{D}\int_0^1{\rm d}\tilde{r}\tilde{r}^{|l|+1}P_j^{(|l|,0)}(1-2\tilde{r}^2)\phiext_l(\tilde{r}) ,
\end{equation}
defined in terms of $\phiext_l(r) = (2\pi)^{-1}\int_0^{2\pi}{\rm d}\varphi\Phi^{\rm ext}(\Rb)\ee^{-\ii l\varphi}$.
 
For a point dipole with moment $\pb$ located a distance $z_0$ above the center of a graphene disk, the elements of ${\bf b}^{(l)}$ are found as \cite{ramirez2015extreme,karanikolas2016tunable}
\begin{equation}
    b_{j,\gamma}^{(l)} = \frac{2\epseff}{\pi\eps_1 D^3} \int_0^1 {\rm d}\tilde{r} \frac{\tilde{r}p_{j,\gamma}^{(l)}(\tilde{r},\tilde{z}_0)}{\left(\tilde{r}^2 + \tilde{z}_0^2\right)^{3/2}} ,
\end{equation}
where $\tilde{z}_0=2z_0/D$ and $\gamma=\{\parallel,\perp\}$ indicates parallel ($\parallel$) or perpendicular ($\perp$) dipole orientations relative to the plane of the disk, distinguished by the elements
\begin{subequations}
\begin{align}
    p_{j,\parallel}^{(l)}(\tilde{r},\tilde{z}_0) &= p\tilde{r}^2 P_j^{(1,0)}(1-2\tilde{r}^2)\left(\delta_{l,1} + \delta_{l,-1}\right) , \\
    p_{j,\perp}^{(l)}(\tilde{r},\tilde{z}_0) &= -2p\tilde{z}_0 P_j^{(0,0)}(1-2\tilde{r}^2)\delta_{l,0} .
\end{align}
\end{subequations}
By solving for the coefficients $a_j^{(l)}$ in Eq.~\eqref{eq:General_matrixEOM_disk}, and thereby for the charge density given by Eq.~\eqref{eq:rho_l_Jacobi_expansion}, we can determine the induced potential entering Eq.~\eqref{eq:Phiind_rho}, and extract the Green's tensor from the relations $\Eb^{\rm ind}(\rb_0,\ww) = -\nabla_{\rb_0}\Phi(\rb_0,\ww) = \ww^2 \mu_0 \Gm^{\rm ref}_{\ww}(\rb_0,\rb_0) \cdot \pb$.

Following the above prescription, we calculate the Purcell factor and Lamb shift in Fig.~\ref{fig4}(b,c) for a dipole located $z=30$\,nm above the center of a $D=200$\,nm graphene disk doped to a Fermi energy is $\EF=0.4$\,eV, with mobility of $\mu_{\rm DC}=5000$\,cm$^2/$Vs at zero temperature. The applied perpendicular magnetic field induces SdH-oscillations in the interband region, as in the previously considered systems, while the near-field response of magnetoplasmon resonances in the disk are found to undergo a spectral splitting proportional to the magnetic field strength. This splitting behavior, which has been observed in far-field measurements of graphene nanodisks illuminated by plane waves \cite{crassee2012intrinsic,wang2012edge,eriksen2025chiral}, also manifests in the dissymmetry factor plotted in Fig.~\ref{fig4}(d), which exhibits a strong chiral character for the dipolar plasmonic resonances. Notably, the SdH oscillations in the interband spectral regime also exhibit near-field dissymmetry, albeit weaker than that associated with magnetoplasmons. The dissymmetry values reached for the geometry under consideration at $B=5$\,T ($B=2.5$\,T) reach $|g|\approx 1.82$ ($|g|\approx 1.46$), and are significantly larger than those achieved for dipole above a graphene ribbon with similar parameters. Fig.~\ref{fig4}(e) similarly reveals a splitting in the Lamb shift for LCP versus RCP dipoles located above the magnetooptical graphene nanodisks, where the splitting ensures a large difference in the Lamb shift for a LCP versus RCP dipole under moderate magnetic field amplitudes. The chiral control and emission of dipoles above graphene is therefore enhanced when reducing the graphene dimensionality, i.e., we find larger dissymmetry factors moving from 2D to 1D configurations, and again from 1D to 0D. 

\section{Conclusions}

We present a comprehensive investigation of the electromagnetic near-fields provided by the proportionately large magneto-optical response of graphene. Our findings indicate that magneto-optical graphene constitutes a versatile platform to actively control quantum light generation, with both the free charge carrier density and the strength of the applied magnetic field offering unprecedented tunability over the photon generation rate and the spectral properties of emitted photons. In extended graphene, magnetoplasmons provide large Purcell factors and Lamb shifts at low energies, but are less tunable with respect to magnetic field and carrier density. In contrast, Shubnikov-de-Haas oscillations induced by moderate static magnetic fields provide sharp spectral variations in the local photonic density of states at energies $\hbar\omega\gtrsim2\EF$, corresponding to near-IR or telecommunications wavelengths at doping levels $\EF\sim0.5$\,eV that are achievable via electrostatic gating. Moreover, our calculations reveal the important role of Landau level degeneracy in the active electrical control of magneto-optical graphene, particularly under cryogenic conditions, where the discrete energy spectrum of electronic states can produce discontinuities in the spectral response as the Fermi energy is continuously varied and a reservoir of charge carriers is available. However, these spectral jumps to not appear if the charge carrier density is conserved in a closed system.

Simulations of optical extinction from graphene nanoribbons reveal qualitatively similar behavior as in extended samples, with localized magnetoplasmon resonances sustained by these low-dimensional systems offering a greater degree of electrical and magnetic tunability owing to lateral spatial confinement. The quantum electrodynamic response of an emitter interfacing magneto-optical graphene nanoribbons is similarly dominated by magnetoplasmons at low energies and SdH oscillations in the interband electronic transition regime. Interestingly, the magneto-optical graphene nanoribbon is shown to produce a large chiral response, quantified by substantial dissymmetry in the emission from left-hand and right-hand circularly polarized dipoles, which is at odds with the negligible chiral dissymmetry predicted for an extended graphene sheet. These results indicate that the geometrical shaping of near-fields in patterned graphene enable access to chiral light-matter interactions at the nanoscale. In a \textit{lower}-dimensional system, represented by a graphene nanodisk, simulations reveal much greater dissymmetry factors for circularly polarized dipoles, indicating that magnetoplasmons in nanopatterned graphene can enable strong chiral control of quantum emitters.

Our findings are based on versatile semianalytical formalisms that we consolidate here to describe nanoscale light-matter interactions in gyrotropic 2D materials. While we have focused here on graphene, the framework we introduce here to describe the near-field response of nanoribbons and nanodisks can be generally applied to other 2D materials characterized by a gyrotropic conductivity tensor. We envision that our findings could inspire experimental investigations of quantum light generation in actively tunable magneto-optical nanophotonic environments.

\section{Acknowledgements}

We thank N.~M.~R.~Peres for stimulating discussions and C.~Tserkezis for proofreading the manuscript.
M.~H.~E. and J.~D.~C. acknowledge support from Independent Research Fund Denmark (grant no. 0165-00051B).
The Center for Polariton-driven Light--Matter Interactions (POLIMA) is funded by the Danish National Research Foundation (Project No.~DNRF165).

\section{Author contributions}

J.~D.~C. proposed the study. M.~H.~E. and J.~D.~C. developed the theory, discussed the results, and wrote the manuscript. M.~H.~E. performed the numerical calculations.

\appendix
\section{Optical response of a gyrotropic two-dimensional material}
In this section we present a derivation of the Fresnel coefficients characterizing the reflection and transmission of electromagnetic waves impinging on a gyrotropic two-dimensional (2D) material, here assumed to lie within the $x$-$y$ plane at $z=0$ that interfaces homogeneous isotropic dielectric media with permittivity $\epsilon_1$ for $z>0$ and $\epsilon_2$ for $z<0$. The obtained Fresnel coefficients are then used to calculate the Green's dyadic for a dipole situated a finite distance $z$ above the 2D material. The derivations presented here are based on the approach of Ref.~\cite{hohenester2020nano}, but generalized to treat gyrotropic media.

\subsection{Fresnel coefficients} \label{app:Fresnel}

In general, the electromagnetic boundary conditions that relate the electric fields $\Eb_{\rm m}$ and the magnetic field $\Bb_{\rm m}$ at a planar interface in the $z=0$ plane that separates regions ${\rm m}\in\{1,2\}$ comprised of otherwise homogeneous media characterized by isotropic permittivity $\epsilon_{\rm m}$ and permeability $\mu_{\rm m}$ are
\begin{subequations}
    \begin{align}
        \hat{\nb}\cdot \left.\ccpar{\Db_1-\Db_2}\right\vert_{z=0} &= \rho^{\rm ext} , \label{eq:BC1} \\
        \hat{\nb}\cdot \left.\ccpar{\Bb_1-\Bb_2}\right\vert_{z=0} &= 0 , \label{eq:BC2} \\
        \hat{\nb}\times\left.\ccpar{\Eb_1-\Eb_2}\right\vert_{z=0} &= 0 , \label{eq:BC3} \\
        \hat{\nb}\times\left.\ccpar{\Hb_1-\Hb_2}\right\vert_{z=0} &= \Kb^{\rm ext} , \label{eq:BC4} 
    \end{align}
\end{subequations}
where $\Db_{\rm m}=\epsilon_{\rm m}\Eb_{\rm m}$ is the electric displacement field, $\Hb_{\rm m}=\Bb_{\rm m}/\mu_{\rm m}$ is the magnetizing field, and $\hat{\nb}$ is the normal vector pointing from medium ${\rm m}=2$ to medium ${\rm m}=1$, while the boundary conditions account for the presence of external surface charge and current densities $\rho^{\rm ext}$ and $\Kb^{\rm ext}$, respectively, at the interface. To consider a 2D material at the interface, we introduce the external current density
\begin{gather} \label{eq:Kext}
    \Kb^{\rm ext}=\begin{pmatrix}
        \sigma_{xx} & \sigma_{xy} & 0 \\ \sigma_{yx} & \sigma_{yy} & 0 \\ 0 & 0 & 0
    \end{pmatrix} \left.\begin{pmatrix}
        E_x \\ E_y \\ E_z
    \end{pmatrix}\right\vert_{z=0} ,
\end{gather}
expressed in terms of the elements of the 2D conductivity tensor $\sigma$, which we assume has non-vanishing off-diagonal components. Eqs.~\eqref{eq:BC3} and \eqref{eq:BC4} then yield
\begin{subequations} \label{eq:BC_with_gyrotropic_cond}
    \begin{align}
        \left.\ccpar{E_{1x}-E_{2x}}\right\vert_{z=0} &= 0 , \\
        \left.\ccpar{E_{1y}-E_{2y}}\right\vert_{z=0} &= 0 , \\
        \left.\ccpar{H_{1x}-H_{2x}}\right\vert_{z=0} &= \left.\ccpar{\sigma_{yx}E_x+\sigma_{yy}E_y}\right\vert_{z=0} , \\
        \left.\ccpar{H_{1y}-H_{2y}}\right\vert_{z=0} &= -\left.\ccpar{\sigma_{xx}E_x+\sigma_{xy}E_y}\right\vert_{z=0} ,
    \end{align}
\end{subequations}
from which we can compute the Fresnel coefficients for transverse electric (TE) and transverse magnetic (TM) polarized fields.

\subsubsection{TE polarization}

For TE polarized electromagnetic waves oscillating with frequency $\omega$, we consider electric fields fully polarized in the $\yy$ direction according to the ansatz
\begin{subequations}
\begin{align}
    E_{1y} &= E_0 \ccpar{\ee^{-\ii k_{1z} z} + r_{\rm ss}\ee^{\ii k_{1z} z}} , \\
    E_{2y} &= E_0 t_{\rm ss}\ee^{-\ii k_{2z} z} ,
\end{align}
\end{subequations}
where $E_0$ denotes the amplitude of the electric field, $k_{{\rm m}z}=\sqrt{\eps_{\rm m}\mu_{\rm m}\ww^2-k_\parallel}$ denotes the normal component of the optical wave vector $\kb_{\rm m}$ in medium ${\rm m}$, expressed in terms of the conserved in-plane wave vector component $k_\parallel=\sqrt{k_x^2+k_y^2}$, and we have introduced the reflection and transmission coefficients $r_{\rm ss}$ and $t_{\rm ss}$ that relate incoming and outgoing ``senkrecht-'' or s-polarized TE fields. Invoking Faraday's law in the frequency domain, $\nabla\times\Eb=\ii\ww\mu\Hb$, we obtain the corresponding magnetic field
\begin{subequations}
    \begin{align}
        H_{1x} &= \frac{k_{1z}E_0}{\ww\mu_1}\ccpar{\ee^{-\ii k_{1z} z} - r_{\rm ss}\ee^{\ii k_{1z} z}} ,  \\
        H_{2x} &= \frac{k_{2z}E_0}{\ww\mu_2}t_{\rm ss}\ee^{-\ii k_{2z} z} .
    \end{align}
\end{subequations}
Now, to account for the generation of ``parallel'' or p-polarized TM fields, i.e., where the associated magnetic fields are entirely polarized in $\yy$, by s-polarized electromagnetic waves due to the off-diagonal 2D conductivity tensor components in Eq.~\eqref{eq:Kext}, we introduce the cross-polarization Fresnel coefficients $r_{\rm ps}$ and $t_{\rm ps}$ according to
\begin{subequations}
    \begin{align}
        H_{1y} &= Z_1^{-1} E_0 r_{\rm ps} \ee^{ \ii k_{1z} z} , \\
        H_{2y} &= Z_2^{-1} E_0 t_{\rm ps} \ee^{-\ii k_{2z} z} ,
    \end{align}
\end{subequations}
where $Z_{\rm m}=\sqrt{\mu_{\rm m}/\eps_{\rm m}}$ denotes the impedance in medium ${\rm m}$. From Ampere's law, $\nabla\times\Hb=-\ii\ww\eps\Eb$, the corresponding electric fields are
\begin{subequations}
    \begin{align}
        E_{1x} &=  \frac{k_{1z}}{Z_1\ww\eps_1} E_0 r_{\rm ps} \ee^{ \ii k_{1z} z} , \\
        E_{2x} &= -\frac{k_{2z}}{Z_2\ww\eps_2} E_0 t_{\rm ps} \ee^{-\ii k_{2z} z} .
    \end{align}
\end{subequations}
Introducing the above cross-polarized fields in the boundary conditions of Eq.~\eqref{eq:BC_with_gyrotropic_cond} we obtain
\begin{subequations}
\begin{align}
    1+r_{\rm ss}-t_{\rm ss} &= 0 , \\
    \frac{k_{1z}}{\eps_1 Z_1}r_{\rm ps} + \frac{k_{2z}}{\eps_2 Z_2} t_{\rm ps} &=0 , \\
    \frac{1}{Z_1}r_{\rm ps} - \frac{1}{Z_2}t_{\rm ps} &= -\sigma_{xx} \frac{k_{1z}}{\ww \eps_1 Z_1}r_{\rm ps} - \sigma_{xy} (1+r_{\rm ss}) , \\
    \frac{k_{2z}}{\ww \mu_2} t_{\rm ss} - \frac{k_{1z}}{\ww \mu_1}(1-r_{\rm ss})  &= -\sigma_{yx} \frac{k_{1z}}{\ww \eps_1 Z_1} r_{\rm ps} - \sigma_{yy} (1+r_{\rm ss}) .
\end{align}
\end{subequations}
Solving this system of linear equations using, e.g., Cramer's rule, we isolate the reflection and transmission coefficients
\begin{subequations}
\begin{align}
    r_{\rm ss} &= \frac{\Dm_{\eps}^{(+)} \Dm_{\mu}^{(-)} + \sigma_{xy}\sigma_{yx}}{\Dm_{\eps}^{(+)} \Dm_{\mu}^{(+)} - \sigma_{xy}\sigma_{yx}} , \\
    t_{\rm ss} &= \frac{2}{\mu_1} \frac{\Dm_{\eps}^{(+)} k_{1z}}{\Dm_{\eps}^{(+)} \Dm_{\mu}^{(+)} -\sigma_{xy}\sigma_{yx}} , \\
    r_{\rm ps} & = - \frac{k_{2z}k_1}{k_{1z}k_2} t_{\rm ps}= -\frac{2}{Z_1} \frac{\sigma_{xy}}{\Dm_{\eps}^{(+)} \Dm_{\mu}^{(+)} - \sigma_{xy}\sigma_{yx}} , 
\end{align}
\end{subequations}
where we have defined
\begin{subequations}
\begin{align}
    \Dm_{\eps}^{(\pm)} &\equiv \frac{\eps_1}{k_{1z}} \pm \frac{\eps_2}{k_{2z}} \pm \frac{\sigma_{xx}}{\ww} , \\
    \Dm_{\mu}^{(\pm)} &\equiv \frac{k_{1z}}{\mu_1} \pm \frac{k_{2z}}{\mu_2} \pm \ww \sigma_{yy}
\end{align}
\end{subequations}
to simplify the expressions.

\subsubsection{TM polarization}

The case of TM polarized electromagnetic waves follows in an entirely analogous manner from the derivation for TE polarization by using the ansatz
\begin{subequations}
\begin{align}
    H_{1y} &= Z_1^{-1} E_0 \ccpar{\ee^{-\ii k_{1z} z} + r_{\rm pp} \ee^{\ii k_{1z} z}} , \\
    H_{2y} &= Z_2^{-1} E_0 t_{\rm pp} \ee^{-\ii k_{2z} z} , \\
    E_{1y} &= E_0 r_{\rm sp} \ee^{\ii k_{1z} z} , \\
    E_{2y} &= E_0 t_{\rm sp} \ee^{-\ii k_{2z} z} ,
\end{align}
\end{subequations}
which leads to the Fresnel coefficients
\begin{subequations}
\begin{align}
    r_{\rm pp} &= -\frac{\Dm_{\eps}^{(-)}\Dm_{\mu}^{(+)} + \sigma_{xy}\sigma_{yx}}{\Dm_{\eps}^{(+)}\Dm_{\mu}^{(+)} - \sigma_{xy}\sigma_{yx}} , \\
    t_{\rm pp} &= \frac{2}{Z_1} \frac{\Dm_{\mu}^{(+)}/\ww}{\Dm_{\eps}^{(+)}\Dm_{\mu}^{(+)} -\sigma_{xy}\sigma_{yx}} , \\
    r_{\rm sp} &= t_{\rm sp} = \frac{2}{Z_1} \frac{\sigma_{yx}}{\Dm_{\eps}^{(+)}\Dm_{\mu}^{(+)} -\sigma_{xy}\sigma_{yx}} .
\end{align}
\end{subequations}
The Fresnel coefficients above for TE and TM polarization are in agreement with those reported in Ref.~\cite{kortkamp2015active}.

\subsection{Green's dyadic} \label{app:Greens}

The electric field produced at $\rb$ in a medium with permeability $\mu$ by a point dipole with moment $\pb$, oscillating at frequency $\omega$, and located at $\rb'$ can be expressed as
\begin{equation}
    \Eb(\rb) = \mu \ww^2 \left[ \Gm_\ww^{0}(\rb,\rb') + \Gm_\ww^{\rm ref}(\rb,\rb') \right] \cdot \pb ,
\end{equation}
where $\Gm_\ww^{(0)}$ and $\Gm_\ww^{\rm ref}$ denote the bare and reflected parts, respectively, of the classical dyadic Green's tensor $\Gm_\ww=\Gm_\ww^{(0)}+\Gm_\ww^{\rm ref}$. Following Ref.~\cite{hohenester2020nano}, we evaluate the Green's tensor for fields in a medium with isotropic permittivity $\epsilon_1$ and permeability $\mu_1$ above a planar interface in the $z=0$ plane characterized by Fresnel reflection coefficients $r_{\alpha\beta}$ associated with incoming (outgoing) $\alpha$-polarized ($\beta$-polarized) light for $\alpha,\beta\in\{{\rm s},{\rm p}\}$ as
\begin{widetext}
\begin{align}
    &\Gm_{\ww,ij}(\rb,\rb') = -\frac{\hat{z}_i \hat{z}_j }{k_1^2} \delta(\rb - \rb') \nonumber \\
    &+\frac{\ii}{8\pi^2}\int_{-\infty}^{\infty}{\rm d}k_x\int_{-\infty}^{\infty}{\rm d}k_y\frac{1}{k_{1z}} \ee^{\ii \sqpar{k_x(x-x')+k_y(y-y')}} \begin{Bmatrix}
        \left[ \ee^{\ii k_{1z} |z-z'|} e_i^{\rm s}(\kb_1^{\pm}) + r_{{\rm ss}}\ee^{\ii k_{1z} (z+z')} e_i^{\rm s}(\kb_1^+) \right] \cdot e_j^{\rm s} (\kb_1^{\pm}) \\
        + \left[ \ee^{\ii k_{1z} |z-z'|} e_i^{\rm p}(\kb_1^{\pm}) + r_{{\rm pp}}\ee^{\ii k_{1z} (z+z')} e_i^{\rm p}(\kb_1^+) \right] \cdot e_j^{\rm p} (\kb_1^{\pm}) \\
        + r_{{\rm ps}}\ee^{\ii k_{1z} (z+z')} e_i^{\rm p}(\kb_1^+) \cdot e_j^{\rm s} (\kb_1^{\pm}) \\
        + r_{{\rm sp}}\ee^{\ii k_{1z} (z+z')} e_i^{\rm s}(\kb_1^+) \cdot e_j^{\rm p} (\kb_1^{\pm})
    \end{Bmatrix} , \label{eq:G_ij}
\end{align}
where
\begin{subequations}
\begin{align}
    \hat{\eb}^{\rm s} (\kb_{\rm m}^{\pm}) &= \frac{\kb_{\rm m}^{\pm}\times \hat{\zb}}{|\kb_{\rm m}^{\pm}\times\hat{\zb}|} = \frac{1}{\sqrt{k_x^2+k_y^2}}(k_y\hat{\xb} - k_x\hat{\yb}) , \\
    \hat{\eb}^{\rm p}(\kb_{\rm m}^{\pm}) &=\kb_{\rm m}^{\pm} \times \hat{\eb}^{\rm s} (\kb_{\rm m}^{\pm}) = \frac{\pm k_{{\rm m}z}}{k_{\rm m} \sqrt{k_x^2+k_y^2}} (k_x\hat{\xb} + k_y\hat{\yb} ) - \frac{1}{k_m}\sqrt{k_x^2 + k_y^2}\hat{\zb}
\end{align}
\end{subequations}
\end{widetext}
are polarization vectors given in terms of $\kb_{\rm m}^{\pm} = k_x \hat{\xb} + k_y \hat{\yb} \pm k_{{\rm m}z} \hat{\zb}$, with the sign of $k_{{\rm m}z} = \sqrt{k_{\rm m}^2 - k_x^2 - k_y^2 + \ii 0^+}$ corresponding to upgoing ($+$) and downgoing ($-$) waves in media with permittivity $\epsilon_{\rm m}$ and permeability $\mu_{\rm m}$.

The reflected part of the Green's tensor in Eq.~\eqref{eq:G_ij} can be written in the form
\begin{widetext}
\begin{equation} \label{eq:G_kx_ky}
   \Gm_\ww^{\rm ref}(\rb,\rb')=\frac{\ii }{8\pi^2}\int_{-\infty}^{\infty}{\rm d}k_x\int_{-\infty}^{\infty}{\rm d}k_y\ee^{\ii\sqpar{k_x(x-x')+k_y(y-y')+k_{1z}(z+z')}}\ccpar{\mathcal{M}^{\rm ss} + \mathcal{M}^{\rm pp} +\mathcal{M}^{\rm sp} + \mathcal{M}^{\rm ps}} ,
\end{equation}
where
\begin{subequations} \label{eq:M_kx_ky}
\begin{align}
     \mathcal{M}^{\rm ss} &= \frac{r_{{\rm ss}}}{k_{1z}}\hat{\eb}^{\rm s}(\kb_1^+) \otimes  \hat{\eb}^{\rm s}(\kb_1^{\pm}) = \frac{r_{{\rm ss}} }{k_{1z}(k_x^2+k_y^2)} \begin{bmatrix}
         k_y^2 & -k_xk_y&0 \\ -k_xk_y & k_x^2 & 0 \\ 0 & 0 & 0
     \end{bmatrix} , \\
     \mathcal{M}^{\rm pp} &= \frac{r_{{\rm pp}}}{k_{1z}} \hat{\eb}^{\rm p} (\kb_1^+) \otimes \hat{\eb}^{\rm p}(\kb_1^{\pm}) = \frac{r_{{\rm pp}}}{k_1^2(k_x^2+k_y^2)} \begin{bmatrix}
         \pm k_{1z} k_x^2 & \pm k_xk_yk_{1z} & -k_x(k_x^2+k_y^2)\\ \pm k_xk_yk_{1z} & \pm k_{1z}k_y^2 & -k_y(k_x^2 + k_y^2) \\ \mp k_x(k_x^2+k_y^2) & \mp k_y(k_x^2+k_y^2) & (k_x^2+k_y^2)^2/k_{1z}
     \end{bmatrix} , \\     
     \mathcal{M}^{\rm ps} &= \frac{r_{{\rm ps}}}{k_{1z}} \hat{\eb}^{\rm p} (\kb_1^+) \otimes \hat{\eb}^{\rm s}(\kb_1^{\pm}) = \frac{r_{{\rm ps}} }{k_1k_{1z}(k_x^2+k_y^2)} \begin{bmatrix} k_xk_yk_{1z} & -k_x^2k_{1z} & 0 \\
     k_y^2k_{1z} & -k_xk_yk_{1z} & 0 \\
     -k_y(k_x^2+k_y^2) & k_x(k_x^2+k_y^2) & 0
     \end{bmatrix} , \\      
          \mathcal{M}^{\rm sp} &= \frac{r_{{\rm sp}}}{k_{1z}} \hat{\eb}^{\rm s} (\kb_1^+) \otimes \hat{\eb}^{\rm p}(\kb_1^{\pm}) = \frac{r_{{\rm sp}} }{k_1k_{1z}(k_x^2+k_y^2)} \begin{bmatrix} \pm k_{1z} k_yk_x & \pm k_{1z} k_y^2 & -k_y(k_x^2+k_y^2) \\ \mp k_x^2k_{1z} & \mp k_yk_xk_{1z} & k_x(k_x^2+k_y^2)\\ 0 & 0 & 0
     \end{bmatrix} .
\end{align}
\end{subequations}
\end{widetext}
The dipole self-interaction is then described by evaluating the integral of Eq.~\eqref{eq:G_kx_ky} in  cylindrical coordinates for $z=z'$ by choosing the lower sign in Eqs.~\eqref{eq:G_kx_ky}. Assuming that the reflection coefficients are independent of the azimuthal angle, i.e., $r_{\alpha\beta}(k_{\parallel},\phi)\to r_{\alpha\beta}(k_{\parallel})$, the Green's tensor reduces to Eq.~\eqref{eq:G_ref_gyrotropic}.

\section{Near-field response of a one-dimensional ribbon} \label{app:1D}

Transforming the external field in Eq.~\eqref{eq:E_ext_dipole} as $\Eb_q^{\rm ext}(x,\ww)=\int{\rm d}y\Eb^{\rm ext}(\Rb,\ww)\ee^{-\ii q y}$, we find
\begin{equation}
    \Eb_q^{\rm ext}(x,\ww) = \frac{1}{4\pi\eps_1}\tensor{g}_q(x,\rb_0)\cdot\pb ,
\end{equation}
where the tensor $\tensor{g}_q(x,\rb)$ is defined in Eq.~\eqref{eq:g_q}. To determine the Green's dyadic for a ribbon, we compute the induced electric field acting back on the dipole at $\rb_0$ in a medium with permittivity $\eps_1$ and permeability $\mu_1$ according to
\begin{align*}
    \Eb^{\rm ind}(\rb_0,\ww) &= -\frac{1}{4\pi\epseff}\nabla_{\rb_0}\int{\rm d}^2\Rb\frac{\rhoind(\Rb,\ww)}{\abs{\rb_0-\Rb}}  \\
    &= \frac{\ii}{4\pi\epseff\ww}\int{\rm d}x\int\frac{{\rm d}q}{2\pi}\tensor{g}_q^*(x,\rb_0)\cdot\Jb_q(x,\ww) ,
\end{align*}
where we have again interchanged the induced charge and current densities using the continuity equation and integrated by parts. Inserting the current of Eq.~\eqref{eq:J_final} in the above expression, we isolate the reflected part of the Green's dyadic from the definition $\Eb^{\rm ind}(\rb_0,\ww) = \ww^2\mu_1\Gm_\ww^{\rm ref}(\rb_0,\rb_0)\cdot\pb$ as 
\begin{align}
    \Gm^{\rm ref}_{\ww}(\rb_0,\rb_0) &= \frac{\ii}{16\pi^2\epseff\ww k_1^2}\int{\rm d}x \int\frac{{\rm d}q}{2\pi} \tensor{g}_q^*(x,\rb_0) \nonumber \\
    & \cdot \Mm_{q,\ww}(x) \cdot f(x)\sigma(\ww) \cdot \tensor{g}_q(x,\rb_0) .
\end{align}
Equipped with the above Green's tensor, we compute the Purcell factor and Lamb shift experienced by a quantum emitter near a ribbon comprised of a 2D material with an arbitrary intrinsic optical conductivity.

Incidentally, the interaction between a dipole $\pb_1$ located at $\rb_1$ and another dipole $\pb_2$ at $\rb_2$ can be quantified by the Green's dyadic
\begin{align}
    \Gm^{\rm ref}_{\ww}(\rb_2,\rb_1) &= \frac{\ii}{16\pi^2\epseff\ww k_1^2}\int{\rm d}x\int\frac{{\rm d}q}{2\pi}\tensor{g}_q^*(x,\rb_2) \nonumber \\
    &\cdot\Mm_{q,\ww}(x)\cdot f(x)\sigma(\ww)\cdot\tensor{g}_q(x,\rb_1)
\end{align}
that mediates the field produced by $\pb_1$ acting on $\pb_2$. In cases where $x_{\rm a}=x_{\rm b}=x_0$ and $z_{\rm a}=z_{\rm b}=z_0$, the interacting Green's tensor reduces to the its self-interacting counterpart up to a phase factor $\ee^{\ii q(y_2-y_1)}$ in the integrand. The interacting Green's tensor could be straightfowardly used to investigate super- or subradiance for multiple emitters mediated by a ribbon, as has been considered in Ref.~\cite{huidobro2012superradiance} for (isotropic) graphene.

\subsection{Ribbon discretization scheme}

The real-space discretization scheme used to obtain the current in Eq.~\eqref{eq:J_final} is based on the approach outlined in Refs.~\cite{christensen2015kerr,rasmussen2023nonlocal}. Specifically, we discretize the space along $x$ into $L$ equidistant points $\{\theta_l\}$, where $\theta=x/W$ is the normalized coordinate, and include points beyond the ribbon to account for the vanishing current at the boundaries~\cite{christensen2015kerr}. The current is similarly discretized as $\Jb_{q,l}=\Jb_q(\theta_l)$, along with other spatially dependent quantities. Adopting a five-point stencil scheme (resulting in fourth-order accuracy in the step size $h=\theta_{l+1}-\theta_l$), the first- and second-order derivative operators away from the ribbon boundaries are
\begin{subequations}
\begin{gather}
    \partial_x \Jb_{q,l} \approx \frac{1}{12h} \left( \delta_{l-2,l'} - 8\delta_{l-1,l'} + 8\delta_{l+1,l'} - \delta_{l+2,l'} \right)\Jb_{q,l'} , \\
    \partial_x^2 \Jb_{q,l} \approx \frac{1}{12h^2} \begin{pmatrix}
          -\delta_{l-2,l'} + 16\delta_{l-1,l'} - 30\delta_{l,l'}  \\
          +16\delta_{l+1,l'} - \delta_{l+2,l'}
    \end{pmatrix} \Jb_{q,l'} .
\end{gather}
\end{subequations}
At the boundary of the computational domain, the five-point central-difference formulation above is not applicable, and so instead we implement one-sided first- and second-order derivatives
\begin{subequations}
\begin{gather}
    \partial_x \Jb_{q,l} \approx \frac{1}{12h} \begin{pmatrix}
        -25\delta_{l,l'} + 48\delta_{l+1,l'} - 36\delta_{l+2,l'} \\
        + 16\delta_{l+3,l'} - 3\delta_{l+4,l'}
    \end{pmatrix} \Jb_{q,l'} , \\
    \partial_x^2 \Jb_{q,l} \approx \frac{1}{12h^2} \begin{pmatrix}
        35\delta_{l,l'} - 104\delta_{l+1,l'} + 114\delta_{l+2,l'} \\
        - 56\delta_{l+3,l'} + 11\delta_{l+4,l'}
    \end{pmatrix} \Jb_{q,l'} ,
\end{gather}
\end{subequations}
where the first-order derivative is of order four and the second-order derivative is of order three \cite{fornberg1988generation}.

The integral kernel is similarly discretized as
\begin{gather}
    \Vm_q\Jb_q(x_l,\ww) = 2\int{\rm d}x'K_0\ccpar{|q||x_l-x'|}\Jb_q(x',\ww) .
\end{gather}
Assuming a slowly varying $\Jb_q(x,\ww)$ and equidistant discretization $\theta_l$, we decompose the operator $\Vm_q$ as a matrix according to $\Vm_q \Jb_q (x_l,\ww)=\sum_l\Vm_{q,ll'}\Jb_{q,l'}$, with matrix elements
\begin{align*}
    \Vm_{q,ll'} &= 2\int_{\theta_{l'}-h/2}^{\theta_{l'}+h/2} d\theta' K_0\ccpar{|q||\theta_l-\theta'|} \\
    &= \pi \sum_{\tilde{\theta}=\theta_{ll'}\pm h/2}(\pm\tilde{\theta})\begin{bmatrix}
        K_0(|q\tilde{\theta}|)\Lm_{-1}(|q\tilde{\theta}|) \\
        + K_1(|q\tilde{\theta}|)\Lm_0(|q\tilde{\theta}|)
    \end{bmatrix}
\end{align*}
where $\Lm_n$ is the modified Struve function of order $n$ \cite{NIST:DLMF}. Note that in the limit of $q\to0$, the kernel of the operator $\Vm$ can be expanded as
\begin{equation}
    K_0\ccpar{|q||\theta-\theta'|}\sim-\log\ccpar{\abs{\theta-\theta'}}-\log(|q|)+\log(2)-\gamma_{\rm EM} ,
\end{equation}
where $\gamma_{\rm EM}$ is the Euler-Mascheroni constant. Invoking the charge neutrality condition, we find that the $\theta'$-independent terms can be ignored \cite{christensen2015kerr}, leading to
\begin{equation}
    \Vm_{0,ll'} = 2\sum_\pm(\pm)\ccpar{\theta_{ll'}\pm h/2}\ccpar{1-\log\abs{\theta_{ll'}\pm h/2}} .
\end{equation}
In practice, we obtain well-converged results using discretization of $L\sim 250$ points.

\section{Optical response of a gyrotropic disk}\label{app:disk}

Following the prescription of Ref.~\cite{eriksen2025chiral} to compute the polarizability of a gyrotropic nanodisk, we express the 2D charge density in a disk of diameter $D$ in cylindrical coordinates $\Rb=(r,\varphi)$ and expand in Jacobi polynomials $P_j^{(|l|,0)}(r)$ according to \cite{fetter1986magnetoplasmons,christensen2014classical}
\begin{equation} \label{eq:rho_l_Jacobi_expansion}
    \rhoind(\Rb) = \sum_{l=-\infty}^{\infty}\sum_{j=0}^{\infty} a_j^{(l)} \tilde{r}^{|l|} P_j^{(|l|,0)} (1-2\tilde{r}^2) \ee^{\ii l\varphi} ,
\end{equation}
where $\tilde{r}=2r/D$ is the normalized radial coordinate. Inserting the density into the Poisson equation and performing algebraic manipulations, we obtain a matrix equation for the vector ${\bf a}^{(l)}$ containing the expansion coefficients $a_j^{(l)}$ introduced above as
\begin{equation} \label{eq:General_matrixEOM_disk}
    \sqpar{\frac{\ii\epseff\ww D}{\sigma_\parallel}\ccpar{\frac{\sigma_\parallel-\ii\,{\rm sgn}(l)\sigma_\perp}{\sigma_\parallel+\ii\,{\rm sgn}(l)\sigma_\perp}\mathcal{A}^{(l)}+\mathcal{B}^{(l)}}+\mathcal{C}^{(l)}}{\bf a}^{(l)}={\bf b}^{(l)} ,
\end{equation}
where the matrices $\mathcal{A}^{(l)}$, $\mathcal{B}^{(l)}$, and $\mathcal{C}^{(l)}$ are comprised of elements
\begin{subequations}\label{eq:Matrices_for_EOM_disk}
    \begin{align}
        \Am^{(l)}_{jj'} &= \frac{\delta_{0j}\delta_{0j'}}{8|l|(|l|+1)^2} ,  \\
        \Bm^{(l)}_{jj'} &= \frac{1}{8(|l|+2j+1)(|l|+2j'+1)}\begin{pmatrix}
            \frac{\delta_{jj'}+\delta_{j,j'+1}}{|l|+2j} \\
            + \frac{\delta_{jj'}+\delta_{j+1,j'}}{|l|+2j+2}
        \end{pmatrix} ,  \\
        \Cm^{(l)}_{jj'} &= \frac{(-1)^{j-j'+2}}{\pi[4(j-j')^2-1](|l|+j+j'+\tfrac{1}{2})(|l|+j+j'+\tfrac{3}{2})} ,
    \end{align}
\end{subequations}
while the vector ${\bf b}^{(l)}$ contains entries
\begin{equation}
    b_j^{(l)} = \frac{4\epseff}{D}\int_0^1 {\rm d}\tilde{r}\tilde{r}^{|l|+1}P_j^{(|l|,0)}(1-2\tilde{r}^2)\phiext_l(\tilde{r})
\end{equation}
that depend on the quantities
\begin{equation} \label{eq:external_pot_lmode}
    \phi_l^{\rm ext}(r) = \int_0^{2\pi}\frac{{\rm d}\varphi}{2\pi}\Phi^{\rm ext}(\Rb)\ee^{-\ii l\varphi}
\end{equation}
associated with the external potential acting on the points $\Rb=(r,\varphi)$ within the nanodisk. Importantly, the matrix elements given in Eq.~\eqref{eq:Matrices_for_EOM_disk} are only valid for $l\neq 0$, whereas the $l=0$ case omits the indices $j=j'=0$, such that $\Am^{(0)}=0$, while the other vector- and matrix-indices instead start at $j=j'=1$~\cite{fetter1986magnetoplasmons,muniz2020two,eriksen2025chiral}. For the purpose of this paper, we truncate the matrices at $j\sim 200$ that ensures convergence.

Once the expansion coefficients of the charge density are determined, the scalar potential at any point in the environment can be computed from
\begin{equation}
    \Phi^{\rm ind}(\rb) = \frac{1}{4\pi\epseff} \int{\rm d}^2\Rb'\frac{\rhoind(\Rb')}{\abs{\rb-\Rb'}} .
\end{equation}
Expanding the spatial dependence of the Coulomb interaction according to Eq.~\eqref{eq:Coulomb_int_expansion} and the charge density in Eq.~\eqref{eq:rho_l_Jacobi_expansion}, we write the induced potential as $\Phi^{\rm ind}(\rb)=\sum_l \phi_l^{\rm ind}(r,z)\ee^{\ii l \varphi}$ and use the orthogonality condition $\int_0^{2\pi}{\rm d}\varphi\ee^{\ii \varphi(l-l')}=2\pi\delta_{ll'}$ to isolate the quantities
\begin{align*}
    \phi_l^{\rm ind}(r,z) =& \frac{D}{4\epseff} \sum_{j=0}^{\infty} a_j^{(l)} \int_0^{\infty}{\rm d}\tilde{r}'\tilde{r}'^{|l|+1} P_j^{(|l|,0)} (1-2\tilde{r}'^2)  \\ &\times\int_0^\infty{\rm d}k J_{|l|}(k\tilde{r})J_{|l|}(k\tilde{r}')\ee^{-k(\tilde{z}_> - \tilde{z}_<)} ,
\end{align*}
where $\tilde{z}\equiv 2z/D$. Making use of the integral relation
\begin{equation}
    \int_0^1 {\rm d}xx^{|l|+1}J_{|l|}(kx)P_j^{(|l|,0)}(1-2x^2) = \frac{1}{k}J_{|l|+2j+1}(k) ,
\end{equation}
the components of the induced potential reduce to
\begin{align*}
    \phi_l^{\rm ind}(r,z) =& \frac{D}{4\epseff}\sum_{j=0}^{\infty}a_j^{(l)}  \\
    &\times \int_0^\infty \frac{{\rm d}k}{k} J_{\abs{l}}(k\tilde{r})J_{\abs{l}+2j+1}(k)\ee^{-k(\tilde{z}_> - \tilde{z}_<)} . 
\end{align*}
Choosing the potential as that of a point dipole $\pb$ located at $\rb_0$, we then use the electric field $\Eb^{\rm ind} = -\nabla\Phi^{\rm ind}$ induced by the external potential entering Eq.~\eqref{eq:external_pot_lmode} and the definition
\begin{equation} \label{eq:E_ind_Gref}
    \Eb^{\rm ind}(\rb_0) = \ww^2 \mu_0 \Gm_{\ww}^{\rm ref}(\rb_0,\rb_0) \cdot \pb
\end{equation}
to identify the reflected Green's tensor components.

In cylindrical coordinates, we evaluate the induced field 
\begin{align} \label{eq:disk_E_ind}
    &\Eb^{\rm ind}(\rb) =- \frac{1}{4\epseff}\sum_{l,j}a_j^{(l)}\ee^{\ii l\varphi} \nonumber \\ 
    &\times\int_0^{\infty}{\rm d}k J_{|l|+2j+1}(k)\ee^{-k|\tilde{z}|}\begin{Bmatrix}
        \sqpar{J_{|l|-1}(k\tilde{r})-J_{|l|+1}(k\tilde{r})}\hat{\rb}  \\
        +(2\ii l/k\tilde{r})J_{|l|}(k\tilde{r})\hat{\varphi}  \\
        \mp 2 J_{|l|}(k\tilde{r})\zz
    \end{Bmatrix} ,
\end{align}
where we have used the relation $\partial_x J_\nu(x) = \sqpar{J_{\nu-1}(x)-J_{\nu+1}(x)}/2$ for Bessel functions and assumed that the disk lies in the $z=0$ plane, such that the upper and lower signs in the $z$-component of the field correspond to $\tilde{z}>0$ and $\tilde{z}<0$, respectively. We furthermore consider situations where the dipole is centered above the disk at $\tilde{r}=0$, which simplifies evaluation of the Bessel functions, i.e., $J_0(0)=1$, $J_1(0)=J_2(0)=0$. In what follows, we consider the induced fields produced by orienting the dipole normally ($\pb = p\hat{\zb}$) or parallel ($\pb=p\hat{\xb}$) to the disk, which corresponds to the conditions $|l|=1$ or $l=0$, respectively. Using the identity $\int_0^{\infty}{\rm d}xJ_{\nu}(\beta x)\ee^{-\alpha x} = \beta^{-\nu}\left(\sqrt{\alpha^2+\beta^2}-\alpha\right)^{\nu}/\sqrt{\alpha^2 + \beta^2}$ (see e.g., Ref.~\cite{gradshteyn2014table}), the induced electric field in Eq.~\eqref{eq:disk_E_ind} reduces to
\begin{equation}\label{eq:arb_Eind_parallel}
    \Eb^{\rm ind}_\parallel(\rb)= -\frac{1}{4\epseff}\sum_{l,j} \frac{a_j^{(l)} \ee^{\ii l \varphi}}{\sqrt{\tilde{z}^2+1}}\left(\sqrt{\tilde{z}^2 +1 }-\tilde{z}\right)^{2+2j}\rr ,
\end{equation}
and
\begin{equation} \label{eq:arb_Eind_orthogonal}
    \Eb^{\rm ind}_\perp(\rb)=\frac{1}{2\epseff}\sum_{j} \frac{a_j^{(0)}}{\sqrt{\tilde{z}^2+1}} \left(\sqrt{\tilde{z}^2 +1 }-\tilde{z}\right)^{1+2j} \zz ,
\end{equation}
for dipoles oriented parallel or perpendicular to the disk, respectively. Importantly, the Green's tensor component $\Gm_{\ww,xx}^{\rm ref}$ is isolated by taking $\varphi=0$ when calculating the induced field acting back on the dipole at $\rb_0$ in Eq.~\eqref{eq:arb_Eind_parallel}, while the cross-polarized element $\Gm_{\ww,yx}^{\rm ref}$ is similarly found by taking $\varphi=\pi/2$. Combining the above expressions and Eq.~\eqref{eq:E_ind_Gref}, we can then determine the relevant Green's tensor elements following the prescription in the main text to obtain the coefficients $a_j^{(l)}$.


%

\end{document}